\documentclass[prl,noshowpacs,english,superscriptaddress,onecolumn]{revtex4}
\usepackage{amsmath,amssymb,amsfonts,mathrsfs}
\usepackage{amsthm}
\usepackage{CJK}
\usepackage{bm}
\usepackage{babel}
\usepackage{graphicx}
\allowdisplaybreaks
\usepackage{braket}
\usepackage[breaklinks]{hyperref}
\usepackage{color}
\hypersetup{colorlinks=true, linkcolor=blue, citecolor=blue, filecolor=blue, urlcolor=blue}
\usepackage{booktabs}
\usepackage{multirow}
\usepackage{longtable}

\begin{document}

\title{Three-dimensional Dirac Phonons with Inversion Symmetry}

\author{Z. J. Chen}
\affiliation{Department of Physics, South China University of Technology, Guangzhou 510640, P. R. China}
\affiliation{Department of Physics $\&$ Institute for Quantum Science and Engineering, Southern University of Science and Technology,
Shenzhen 518055, P. R. China.}
\affiliation{Guangdong Provincial Key Laboratory of Computational Science and Material Design, Southern University of Science and Technology.}
\author{R. Wang}
\affiliation{Institute for Structure and Function $\&$ Department of physics $\&$ Center for Quantum Materials and Devices, Chongqing University, Chongqing 400044, P. R. China.}
\author{B. W. Xia}
\affiliation{Department of Physics $\&$ Institute for Quantum Science and Engineering, Southern University of Science and Technology,
Shenzhen 518055, P. R. China.}
\affiliation{Guangdong Provincial Key Laboratory of Computational Science and Material Design, Southern University of Science and Technology.}
\author{B. B. Zheng}
\affiliation{Department of Physics $\&$ Institute for Quantum Science and Engineering, Southern University of Science and Technology,
Shenzhen 518055, P. R. China.}
\author{Y. J. Jin}
\affiliation{Department of Physics $\&$ Institute for Quantum Science and Engineering, Southern University of Science and Technology,
Shenzhen 518055, P. R. China.}
\affiliation{Guangdong Provincial Key Laboratory of Computational Science and Material Design, Southern University of Science and Technology.}
\author{Yu-Jun Zhao}
\email[]{zhaoyj@scut.edu.cn}
\affiliation{Department of Physics, South China University of Technology, Guangzhou 510640, P. R. China}
\author{H. Xu}
\email[]{xuh@sustech.edu.cn}
\affiliation{Department of Physics $\&$ Institute for Quantum Science and Engineering, Southern University of Science and Technology,
Shenzhen 518055, P. R. China.}
\affiliation{Guangdong Provincial Key Laboratory of Computational Science and Material Design, Southern University of Science and Technology.}

\begin{abstract}
Dirac semimetals associated with bulk Dirac fermions are well-known in topological electronic systems. In sharp contrast, three-dimensional (3D) Dirac phonons in crystalline solids are still unavailable. Here we perform symmetry arguments and first-principles calculations to systematically investigate 3D Dirac phonons in all space groups with inversion symmetry. The results show that there are two categories of 3D Dirac phonons depending on their protection mechanisms and positions in momentum space. The first category originates from the four-dimensional irreducible representations at the high symmetry points. The second category arises from the phonon branch inversion, and the symmetry guarantees Dirac points to be located along the high symmetry lines. Furthermore, we reveal that non-symmorphic symmetries and the combination of inversion and time-reversal symmetries play essential roles in the emergence of 3D Dirac phonons. Our work not only offers a comprehensive understanding of 3D Dirac phonons but also provides significant guidance for exploring Dirac bosons in both phononic and photonic systems.
\end{abstract}

\pacs{73.20.At, 71.55.Ak, 74.43.-f}

\maketitle


Recently, various topological phases in condensed matter physics become the subject of intense studies in electronic systems. These topological phases are associated with specific symmetries, leading to unique nontrivial surface states. For instance, the time-reversal ($\mathcal{T}$) symmetry guarantees the helical surface states with a Dirac linear dispersion in topological insulators \cite{zhang2009topological,qi2011topological}. Such features have been identified to be robust against non-magnetic perturbations, facilitating potential applications in dissipationless devices \cite{qi2011topological}. Subsequently, the concept of band topology has been introduced to semimetals \cite{wan2011topological}, i.e., topological semimetals, which further extend the classification of topological matter. So far, different types of topological semimetals have been theoretically proposed \cite{wan2011topological,wang2016time,weng2015weyl,wang2012dirac,wang2013three,young2012dirac,wu2018mgta}. In particular, three-dimensional (3D) Dirac semimetals, in which the low-energy excitations around fourfold-degenerate Dirac points linearly disperse along all momentum directions, have attracted much attention as they are regarded as 3D analogs of graphene \cite{wang2012dirac,wang2013three,young2012dirac,wu2018mgta}. Delightfully, 3D Dirac semimetals have already been experimentally confirmed in Na$_3$Bi and Cd$_3$As$_2$ \cite{liu2014stable,neupane2014observation,liu2014discovery,xiong2015evidence}.

\begin{table}[!hbp]
\small
\caption{Space groups and corresponding HSPs that possess Dirac phonons, as well as the key operators that guarantee 4D IRs under the restriction of $\mathcal{T}$.}
\renewcommand\arraystretch{1.8}
\setlength{\tabcolsep}{5.5mm}
\begin{tabular}{cc}
\toprule[0.1em]
\hline
\hline
Space group (HSP) & $\emph{R}_{\alpha(\beta)}$  \ \ \ \\
\hline
73(\emph{W}) \& 142/206/230(\emph{P})	 &     $\{C_{2x}|00\frac{1}{2}\}\;\&\;\{C_{2y}|\frac{1}{2}00\}$ \\
52(\emph{S})                 		     &     $\{C_{2x}|0\frac{1}{2}\frac{1}{2}\}\;\&\;\{C_{2z}|\frac{1}{2}00\}$ \\
54(\emph{U}) \& 54(\emph{R})        	 &     $\{C_{2y}|00\frac{1}{2}\}\;\&\;\{C_{2z}|\frac{1}{2}00\}$ \\
56(\emph{U})                 		     &     $\{C_{2y}|0\frac{1}{2}\frac{1}{2}\}\;\&\;\{C_{2z}|\frac{1}{2}\frac{1}{2}0\}$ \\
60(\emph{T})                 		     &     $\{C_{2x}|\frac{1}{2}\frac{1}{2}0\}\;\&\;\{C_{2y}|00\frac{1}{2}\}$ \\
56(\emph{T}) \& 130/138(\emph{R})   	 &     $\{C_{2x}|\frac{1}{2}0\frac{1}{2}\}\;\&\;\{C_{2z}|\frac{1}{2}\frac{1}{2}0\}$
\\
\hline
\hline
\bottomrule[0.1em]
\end{tabular}
\label{Table}
\end{table}

In comparison with fermionic electrons, bosonic systems possess similar but different properties. On the one hand, bosons do not obey the Pauli exclusion principle. As a result, their topological features are effective in the whole energy range. On the other hand, the spinless Bloch functions in bosonic systems are invariant under an even number of $\mathcal{T}$ operations, i.e., $\mathcal{T}^2=1$ \cite{levine1962note}. Such unique features may supply various fascinating properties and potential applications to the family of topological quantum phases. To date, the research progress in topological bosons mainly focuses on the artificial photonic \cite{lu2016symmetry,sepkhanov2008proposed} and phononic \cite{chen2014accidental,liu2011dirac,li2014double,lu2014dirac} crystals. The studies of topological phonons in crystalline solids are still in infancy \cite{zhang2018double,li2018coexistent,singh2018topological,jin2018ideal,jin2018recipe,miao2018observation,zhang2019phononic}. Especially, to our knowledge, 3D Dirac phonons have not been reported in literature. In analogy to 3D Dirac fermions, a 3D Dirac phonon can be regarded as the overlap of two Weyl phonons with opposite chirality, which is protected by a combination of $\mathcal{T}$ and inversion ($\mathcal{P}$) symmetries, i.e., the $\mathcal{PT}$ symmetry \cite{lu2013weyl,ArmitageWeyl,vafek2014dirac}. Around such phonon Dirac point, the quasiparticle excitations exhibit linear dispersion, which can be described by the massless Dirac equation \cite{mei2012first}.

\begin{table}[t]
\small
\caption{Space groups that host Dirac phonons along the HSLs. The superscripts indicate the dimension of IRs. The symbol ``$\oplus$" represents the direct sum of two sets of IRs combined by the $\mathcal{T}$ symmetry.}
\renewcommand\arraystretch{2.0}
\setlength{\tabcolsep}{4.0mm}
\begin{tabular}{cc}
\toprule[0.1em]
\hline
\hline
Space group (HSL) & Irreducible representations \ \ \ \\
\hline
62(\emph{P}/\emph{E}) \& 55/56/58/59(\emph{Q}) 	    & $\varGamma _1^1\oplus\varGamma _2^1,\varGamma _3^1\oplus\varGamma _4^1$ \\
175/176($\Delta$)               					& $\varGamma _3^1\oplus\varGamma _5^1,\varGamma _4^1\oplus\varGamma _6^1$ \\
191-194($\Delta$)               					& $\varGamma _5^2,\varGamma _6^2$ \\
\hline
\hline
\bottomrule[0.1em]
\end{tabular}
\label{Table}
\end{table}

In this work, we identify that there are 92 space groups with $\mathcal{PT}$ symmetry by screening symmetry conditions. All these space groups are investigated to search for 3D Dirac phonons, which can be classified into two categories. The first category possesses the Dirac points at the high symmetry points (HSPs), and the second category possesses the Dirac points along the high symmetry lines (HSLs). High-throughput calculations were performed to search candidates of 3D Dirac phonons (see the computational methods in the Supplemental Material (SM) \cite{SM}). We identify that Si (\emph{cI}16) and Nb$_3$Te$_3$As are representative candidates for each category, respectively.

To realize 3D Dirac phonons, the fourfold degeneracy of phonon branches is a prerequisite. As it is well-known, a trivial degeneracy is often fragile due to the unavoidable perturbations \cite{von1929no}. Fortunately, the perturbation term can be rigorously forbidden in crystalline solids. In other words, the phonon branch crossings are protected by specific crystal symmetries, forming nontrivial crossing points. In principle, the fourfold degeneracy can be achieved in either of two forms: 1) the essential degeneracy at the HSPs, or 2) the accidental degeneracy along the HSLs. Based on the symmetry analysis and irreducible representations (IRs) in 230 space groups \cite{aroyo2006bilbao}, we respectively investigate these two categories of 3D Dirac phonons, and their symmetry constraints are completely identified.

We first focus on the first category of 3D Dirac phonons at the HSPs. Through checking IRs at the HSPs of 92 centrosymmetric space groups, we reveal 20 HSPs (within 17 space groups) that can host 3D Dirac phonons, as listed in Table S1 of the SM \cite{SM}. It is worth noting that the cases of the presence of Dirac nodal-lines or quadratic dispersion have been excluded (see Tables S2 in the SM  \cite{SM}). Further analysis shows that 12 HSPs (within 10 space groups) only possess four-dimensional (4D) IRs (see Table I), and the minimal symmetry condition for the presence of 4D IRs can be described by
\begin{equation}
{\emph{R}_{\alpha(\beta)}}^2=1, \{\emph{R}_\alpha,\emph{R}_\beta\}=0, \{\emph{R}_{\alpha(\beta)},\mathcal{PT}\}=0,
\end{equation}
where $\emph{R}_{\alpha(\beta)}$  represent the non-symmorphic symmetry operators. Besides these 12 HSPs listed in Table I, another 8 HSPs within 7 space groups given in the bottom panel of Table S1 cannot be explained by Eq. (1), and the corresponding analysis is provided in the SM \cite{SM}.

As a typical example, we illustrate the space group \emph{Ibca} (No. 73), in which 4D IR only emerges at \emph{W} with the wave vector $\textbf{\emph{k}}_\emph{W}=(1/2,1/2,1/2)$. In this case, $\emph{R}_\alpha=\emph{S}_{2x}$ and $\emph{R}_\beta=\emph{S}_{2y}$ are two screw rotations involving half lattice translations, which lead to the coordinate transformation as
\begin{gather}
S_{2x}: (x,y,z)\rightarrow(x,-y,-z+1/2), \notag \\
S_{2y}: (x,y,z)\rightarrow(-x+1/2,y,-z).
\end{gather}
The symmetry transformation leads to
\begin{equation}
\emph{S}_{2x}\emph{S}_{2y}=\textbf{\emph{T}}_\emph{z}\emph{S}_{2y}\emph{S}_{2x},
\end{equation}
where $\textbf{\emph{T}}_\emph{z}$ is a unit lattice translation along the \emph{z} direction. At the \emph{W} point, this translational operation brings a phase factor on Bloch states as $e^{i\textbf{\emph{k}}_\emph{W} \cdot \textbf{\emph{T}}_\emph{z}}=-1$, leading to the anti-commutation relation $\{\emph{S}_{2x},\emph{S}_{2y}\}=0$. The other relations in Eq. (1) can be confirmed by employing the same argument.

To determine the dimension of IRs, we start with two groups of Bloch states
\begin{equation}
A: \{|\varphi\rangle,\mathcal{PT}\emph{S}_{2y}|\varphi\rangle\}, B: \{\emph{S}_{2y}|\varphi\rangle,\mathcal{PT}|\varphi\rangle\},
\end{equation}
where $|\varphi\rangle$ can be chosen as an eigenstate of $\emph{S}_{2x}$ with $\emph{S}_{2x}|\varphi\rangle=\pm|\varphi\rangle$ since ${\emph{S}_{2x}}^2=1$. Then the states in different groups are assigned with opposite $\emph{S}_{2x}$ eigenvalues according to the anti-commutation relations in Eq. (1). This suggests that any state in one group can not be a linear combination of the states in the other group. Then we focus on the states in the same group, e.g., the group $A$. We assume that $|\varphi\rangle$ and $\mathcal{PT}\emph{S}_{2y}|\varphi\rangle$ are linearly dependent, i.e., $|\varphi\rangle=\mu\mathcal{PT}\emph{S}_{2y}|\varphi\rangle$, where $\mu$ is a complex constant. The anti-commutation relation $\{\emph{S}_{2x},\mathcal{PT}\}=0$ gives
\begin{equation}
|\varphi\rangle=\mu\mathcal{PT}\emph{S}_{2y} \cdot \mu\mathcal{PT}\emph{S}_{2y}|\varphi\rangle=-|\mu|^2|\varphi\rangle,
\end{equation}
which means that the solution of $\mu$ does not exist, and thus $|\varphi\rangle$ and $\mathcal{PT}\emph{S}_{2y}|\varphi\rangle$ must be linearly independent. A similar argument is also suitable for the group $B$. As a result, we can conclude that the four states in Eq. (4) are linearly independent and have the same eigenvalue of $\hat{H}$. These four states can always be constructed to four degenerate complete orthonormal sets of $\hat{H}$, and thus the IRs are 4D.

Because the non-symmorphic symmetries may induce a higher degeneracy (i.e., ``bands-sticking-together'' effect) \cite{heine2007group}, the 4D IRs of HSPs are guaranteed to locate at the boundary of the first Brillouin zone (BZ). Besides, due to the $\mathcal{T}$ symmetry, the 4D IRs are the direct sum of a pair of conjugated IRs. For all the HSPs in Table I, the dispersion of phonon branches is linear, forming Dirac phonons. In Fig. 1(a), a representative dispersion of Dirac phonons is plotted to make an intuitive description. Note that the electronic bands of two spin channels degenerate in pairs under the constraint of the $\mathcal{PT}$ symmetry in Dirac semimetals. Nevertheless, spinless phonon branches can individually arise along a general \emph{k}-path, resulting in different behaviors between Dirac phonons and Dirac fermions.

\begin{figure}
	\centering
	\includegraphics[scale=0.5]{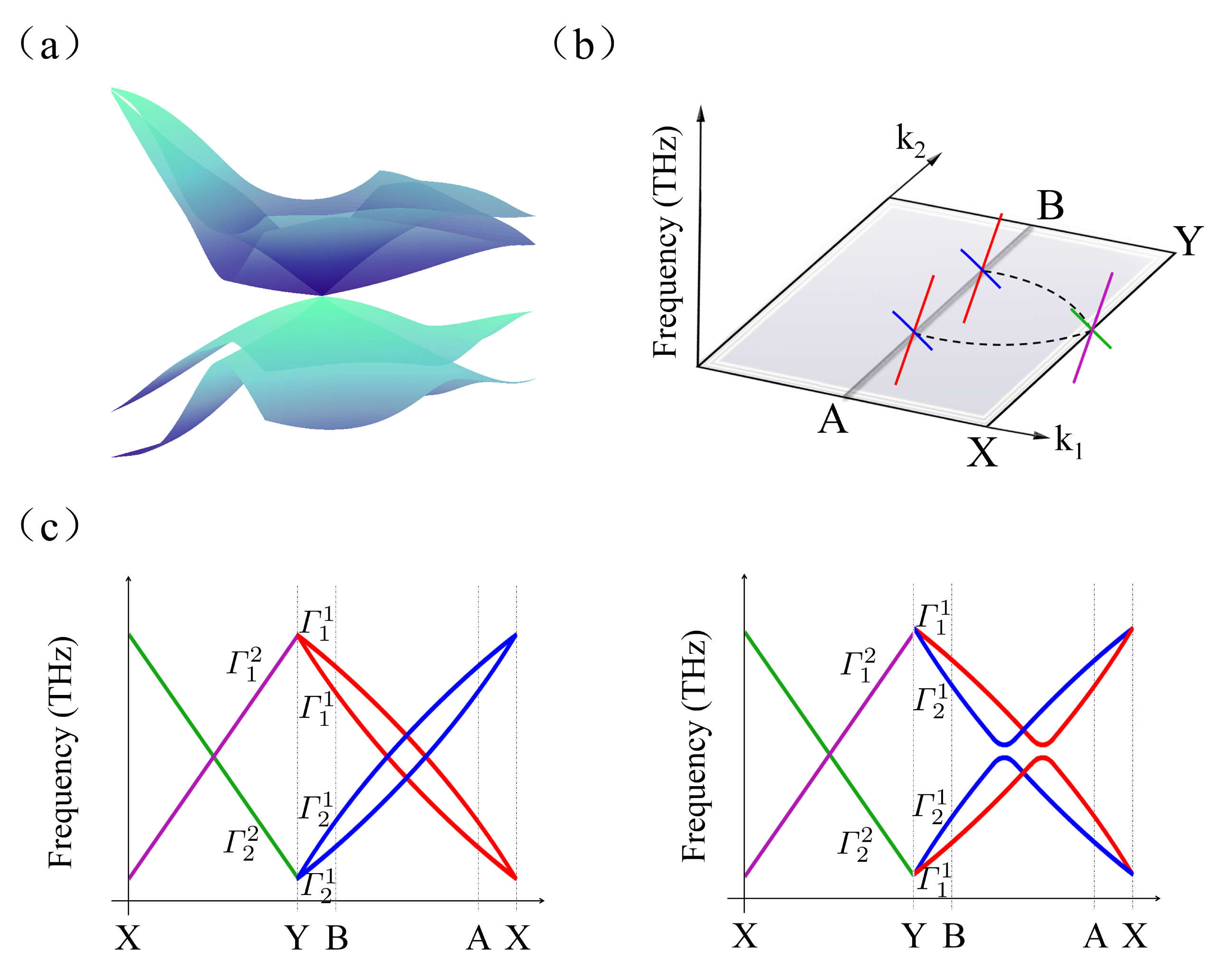}
    \caption{(a) 3D representation of the Dirac phonon. (b) Sketch of the nodal-line induced by accidental degeneracy along the HSL. Red and blue lines indicate $\varGamma_1^1$ and $\varGamma_2^1$, which are 1D IRs in the high symmetry plane. Purple and green lines indicate $\varGamma_1^2$ and $\varGamma_2^2$, which are 2D IRs along the HSL. (c) Two cases of phonon branch splitting along a close path A-B when a symmetry-protected crossing emerges along X-Y.}
\end{figure}

Next, we turn to the second category of Dirac phonons along the HSLs. The Dirac phonons in this category are induced by band inversion. In general, there should be two sets of twofold-degenerate phonon branches crossing each other and they are completely separated away from the crossing point. If these two sets of phonon branches are assigned to different two-dimensional (2D) IRs, the gapless point is symmetry-protected and thus cannot be gapped by local perturbations. Following this rule, we search through all the space groups with inversion symmetry to look for HSLs that possess two or more sets of 2D IRs. The results show that most cases lead to nodal-line phonons in high symmetry planes rather than Dirac phonons along the HSL.

To elaborate on this, we consider the evolution of two sets of 2D IRs $\varGamma_1^2$ and $\varGamma_2^2$, which form a symmetry-protected crossing point along the HSL \emph{X}-\emph{Y} [see Fig. 1]. Besides, we choose an arbitrary path \emph{A}-\emph{B} parallel to \emph{X}-\emph{Y}. These two selected paths lie in a high symmetry plane $k_1$-$k_2$ [see Fig. 1(b)]. When \emph{A}-\emph{B} is infinitely closed to \emph{X}-\emph{Y}, the phonon dispersion along \emph{A}-\emph{B} can be regarded as slight deformations of that along \emph{X}-\emph{Y}. As a result, the 2D IRs $\varGamma_1^2$ and $\varGamma_2^2$ along \emph{X}-\emph{Y} will split into two one-dimensional (1D) IRs along \emph{A}-\emph{B}, i.e., $\varGamma_1^1$ and $\varGamma_2^1$. According to the compatibility relations, there will be two  cases of splitting, either $\varGamma_1^2=\varGamma_1^1\oplus \varGamma_1^1$, $\varGamma_2^2=\varGamma_2^1\oplus \varGamma_2^1$ or $\varGamma_1^2=\varGamma_1^1\oplus \varGamma_2^1$, $\varGamma_2^2=\varGamma_1^1\oplus \varGamma_2^1$. We plot these two cases in Fig. 1(c). In the left panel of Fig. 1(c), the crossings along \emph{A}-\emph{B} will be preserved. This case can happen if the $k_1$-$k_2$ plane is a reflection-invariant plane, and the crossings along \emph{A}-\emph{B} are protected by the mirror symmetry. Then, the continuous deformation of \emph{A}-\emph{B} will generate two nodal-lines in the plane $k_1$-$k_2$ [the black dashed lines in Fig. 1(b)] and the 3D Dirac phonons are not allowed. In the right panel of Fig. 1(c), two branches with the same IR are forbidden to form crossing. In this case, the 3D Dirac phonons can be solely present along \emph{X}-\emph{Y}. In Table S3 of the SM \cite{SM}, we tabulate all the centrosymmetric space groups and HSLs with two or more sets of 2D IRs as well as their compatibility relations with the high symmetry planes. We can see that the 3D Dirac phonons along the HSLs are quite limited. The space groups that can host 3D Dirac phonons along the HSLs are summarized in Table II.

\begin{figure*}[t]
	\centering
	\includegraphics[scale=0.59]{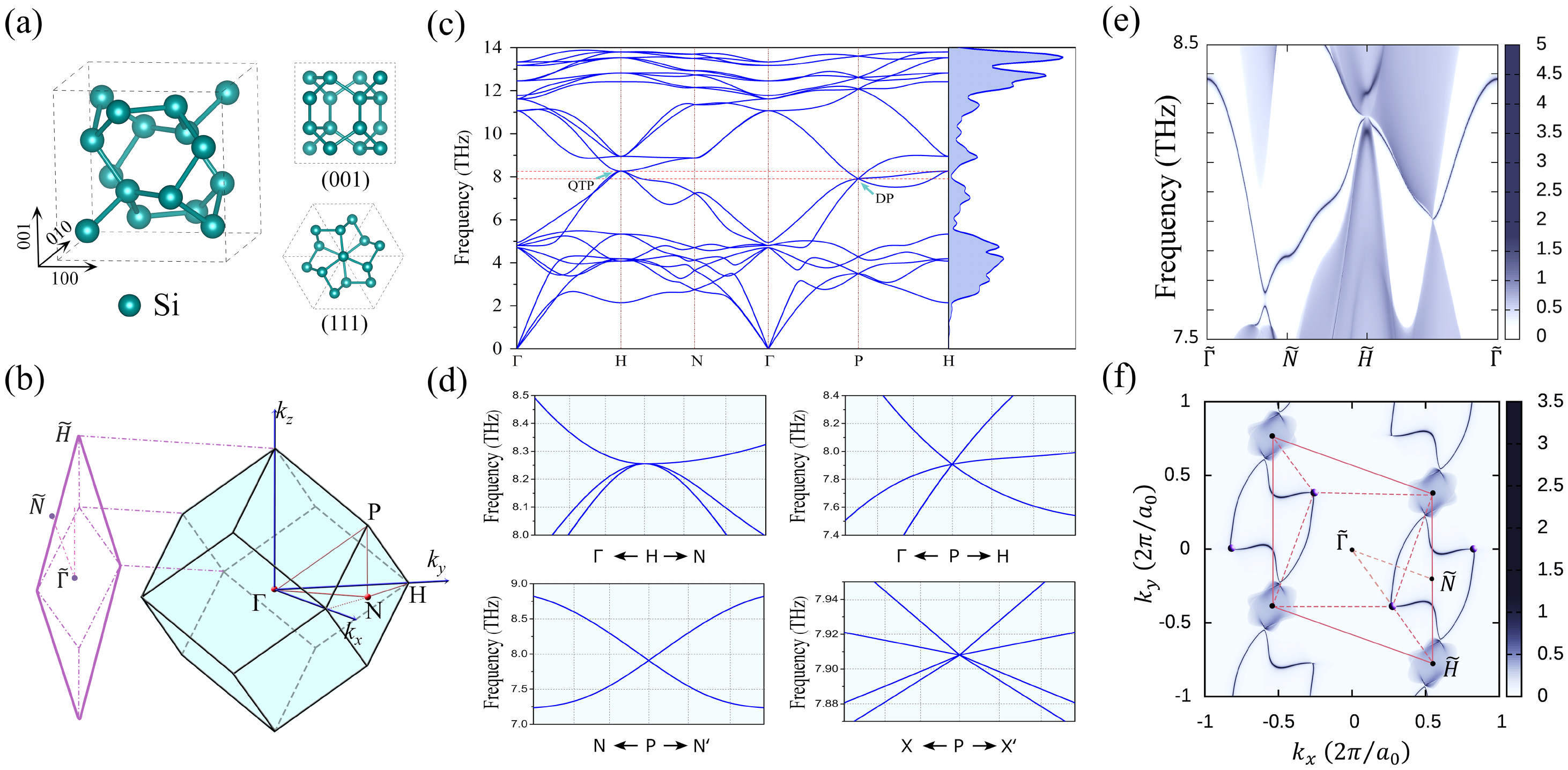}
    \caption{(a) Side and top views of Si (\emph{cI}16) in the conventional cell. (b) The first BZ of the primitive cell and the (110) surface BZ. (c) Phonon spectrum and density of states. (d) Phonon dispersion of QTP and Dirac phonon along different directions. (e) Phonon surface states and (f) arcs projected on the (110) surface.}
\end{figure*}

Based on the above symmetry analysis, we carry out high-throughput screening of phonon-branch topology to search for these two categories of candidates with 3D Dirac phonons. For Dirac points at the HSPs, the existence of independent 4D IRs implies that any material belonging to the proposed space groups could hold such 3D Dirac phonons. In comparison with the case of Dirac points at the HSPs, the search for 3D Dirac phonons along the HSLs is more difficult since there are only a few space groups that satisfy the symmetry conditions. Besides, it is worth noting that the 3D Dirac phonons may usually be hidden in rambling branches, and thus their topological features are invisible. Fortunately, we find several candidates with visible Dirac phonons. Here we take Si (\emph{cI}16) and Nb$_3$Te$_3$As as examples to show 3D Dirac phonons at the HSPs and along the HSLs, respectively. Both materials have been successfully synthesized \cite{wosylus2009crystal,bensch1995preparation}, indicating the feasibility in experiments. Other candidates with 3D Dirac phonons are provided in the SM \cite{SM}.

Si (\emph{cI}16) is a silicon allotrope with space group \emph{Ia}-3 (No. 206). It crystallizes in a body-centered cubic structure with 16 atoms in its primitive unit cell, as shown in Fig. 2(a). The first BZ along with the projected (110) surface BZ are given in Fig. 2(b). We first elucidate that there are 4D IRs at \emph{P} that is invariant under the operations of $\emph{C}_{3,111}$ and $\emph{S}_{2\alpha}=\{C_{2\alpha}|\textbf{\emph{t}}_\beta\}  (\alpha=x,y,z ;\beta=z,x,y)$, where $\emph{C}_{3,111}$ is the threefold rotation along the [111] direction and $\emph{S}_{2\alpha}$ are the twofold screw rotations along the $\alpha$ direction with half lattice translations $\textbf{\emph{t}}_\beta$ along the $\beta$ direction. The little group of \emph{P} satisfies the minimal symmetry condition of Dirac points at the HSPs in Eq. (1) (see the details in the SM \cite{SM}). Ignoring the $\mathcal{T}$ symmetry, we can use the eigenvalues of $\emph{C}_{3,111}$ to represent the IRs as $\emph{P}_1$: \textit{diag}(1, $e^{{i2\pi}/{3}}$), $\emph{P}_2$: \textit{diag}($e^{{i2\pi}/{3}}$, $e^{{-i2\pi}/{3}}$) and $\emph{P}_3$: \textit{diag}($e^{{-i2\pi}/{3}}$, 1). In a phonon system, the complex IRs always appear in pairs as $\mathcal{T}$ is always conserved. As a consequence, the IRs at \emph{P} are given as $P'=P_1\oplus P_3$ and $P''=P_2\oplus P_2$, which are both 4D.

The phonon spectrum of Si (\emph{cI}16) along the high symmetry path is shown in Fig. 2(c). As expected, fourfold degeneracies are present at \emph{P} for all branch nodes. Especially, the linear excitations of Dirac phonons near the frequency of 8 THz are well-separated with other phonon branches, facilitating their detection in experiments. To intuitively show the topological features of 3D Dirac phonons in Si (\emph{cI}16), the enlarged views around the Dirac point along several typical paths are shown in Fig. 2(d). We can find that the degenerate behaviors vary along different directions in momentum space. In addition, it is worth noting that there is a quadratic triple degenerate point (QTP) at \emph{H}, of which the branches are decoupled along the $\varGamma$-\emph{H}-\emph{N} path [see Fig. 2(d)], exhibiting the quadratic dispersion.

\begin{figure}
	\centering
	\includegraphics[scale=0.5]{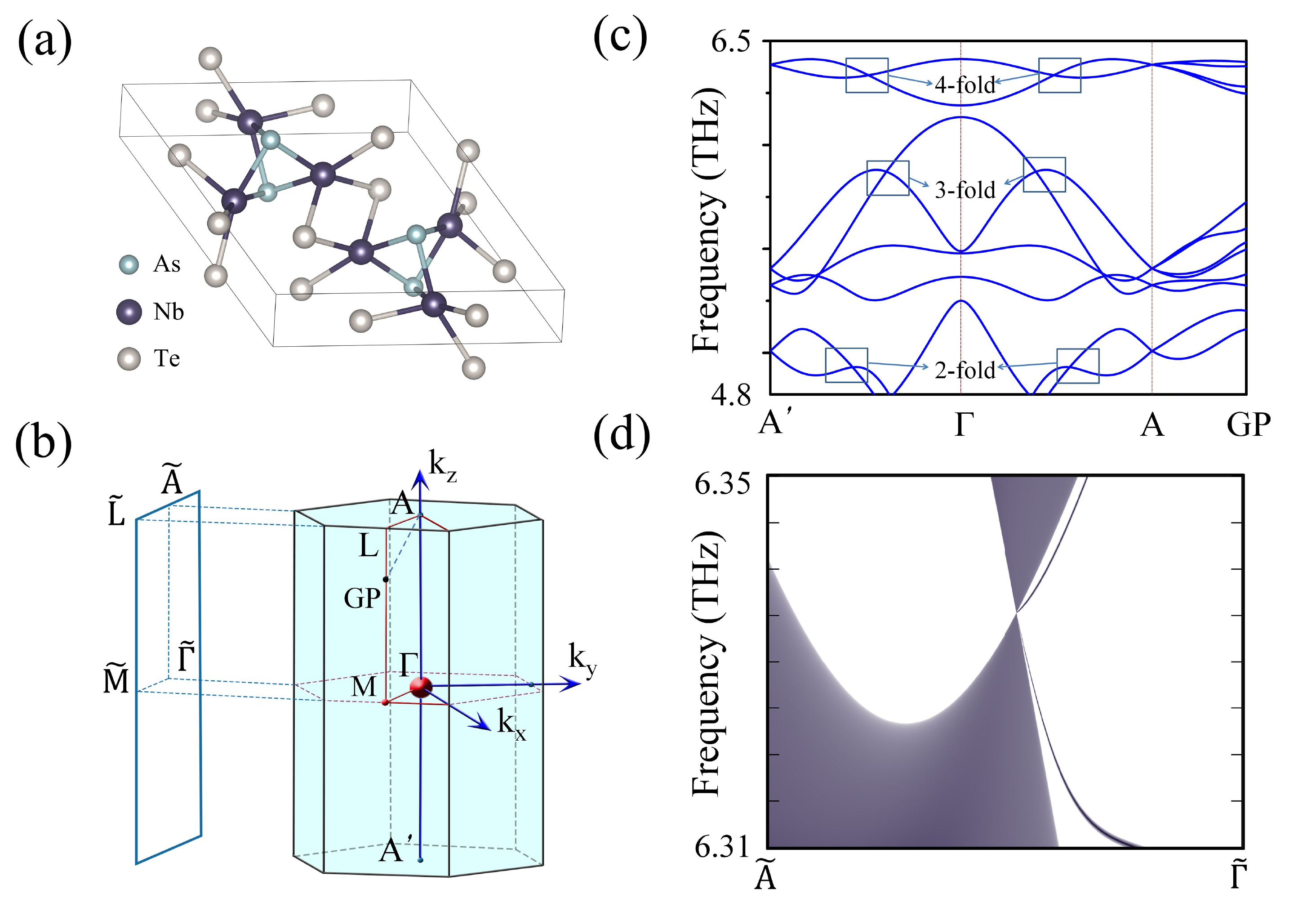}
    \caption{(a) Side view of Nb$_3$Te$_3$As. (b) The first BZ and projected surface BZ parallel to the \emph{k}$_z$ axis. (c) Phonon spectrum of Nb$_3$Te$_3$As ranging from 4.8 THz to 6.5 THz. Square boxes indicate the accidental degeneracies along the \emph{A}'-$\varGamma$-\emph{A} path. (d) Phonon surface states near the Dirac phonon.}
\end{figure}

To obtain topological surface states of Si (\emph{cI}16) at the HSPs, we construct a phonon Wannier tight-binding Hamiltonian using the real-space force constants \cite{wu2018wanniertools}. The calculated phonon local density of states (LDOS) and the corresponding isofrequency surface projected on the (110) surface of Si (\emph{cI}16) are illustrated in Figs. 2(e) and 2(f), respectively. The surface states are composed of two sets. Each set can be viewed as the surface phonon states of Weyl phonons since the Dirac points can be treated as the overlap of two Weyl points with opposite chirality. Figure 2(f) clearly shows that the phonon surface arcs cross over the boundary of the first BZ and connect the projections of two nonequivalent Dirac points.

In the following, we show 3D Dirac phonons in Nb$_3$Te$_3$As along the HSLs. Nb$_3$Te$_3$As crystallizes in a hexagonal structure with space group \emph{P}63/\emph{m} (No. 176), as shown in Fig. 3(a). The bulk BZ and (101) surface BZ are shown in Fig. 3(b). The results show that two phonon branches cross along the \emph{A}'-$\varGamma$-\emph{A} path at the frequency around 6.34 THz [see Fig. 3(c)]. Such crossing points are protected by the screw rotation symmetry \emph{S}$_{6z}$. There are six 1D IRs, which associate with the eigenvalues of \emph{S}$_{6z}$ as $E_n=e^{i\pi n/3}\cdot e^{-ik_zc/2}$, where $c$ is the lattice constant and $n$ is an integer ($n\in$[0,5]). Due to the $\mathcal{T}$ symmetry, (\emph{E}$_1$, \emph{E}$_5$) and (\emph{E}$_2$, \emph{E}$_4$) respectively become two pairs of complex conjugate representations, leading to two sets of twofold-degenerate phonon branches along \emph{A}'-$\varGamma$-\emph{A}. As a result, the Dirac point arising from the crossing between (\emph{E}$_1$, \emph{E}$_5$) and (\emph{E}$_2$, \emph{E}$_4$) is present. Besides, twofold and threefold degenerate points can also be found as 1D IR is allowed on the \emph{k}$_z$ axis. Figure 3(d) gives the phonon surface states along the projected $\tilde{\varGamma}$-$\tilde{A}$ path. There are two branches of nontrivial phonon surface states, which are both terminated at the projected Dirac point, exhibiting the unique topological feature.

To summarize, we have investigated the symmetry conditions for the presence of 3D Dirac phonons in systems with inversion symmetry, and two categories of 3D Dirac phonons (i.e., at the HSPs and along the HSLs) are uncovered. The symmetry arguments reveal that the $\mathcal{PT}$ symmetry and non-symmorphic symmetries play critical roles in the topological classification of 3D Dirac phonons. As a result, all the centrosymmetric space groups that can host 3D Dirac phonons are identified. Furthermore, we provide several realistic materials that realize 3D Dirac phonons. These candidates are expected to be confirmed by experiments, such as inelastic neutron scattering, inelastic X-ray scattering, and He atom scattering. Our findings not only offer the complete topological classification of 3D Dirac phonons in crystalline solids, but can also be extended to 3D Dirac bosons in both phononic and photonic systems.

This work is supported by the Guangdong Natural Science Funds for Distinguished Young Scholars (2017B030306008), the National Natural Science Foundation of China (11974160, 11574088, and 11974062), the Guangdong Provincial Key Laboratory of Computational Science and Material Design (2019B030301001), and the Center for Computational Science and Engineering of Southern University of Science and Technology.

\clearpage

\section*{\large Supplemental Material for ``Three-dimensional Dirac Phonons with Inversion Symmetry"}

\section{Computational methods}
All the calculations were based on the framework of density functional theory (DFT) \cite{kohn1965self} using the Vienna \emph{ab-initio} Simulation Package (VASP) \cite{kresse1996efficient,kresse1996efficiency}. The generalized gradient approximation (GGA) with the Perdew-Burke-Ernzerhof (PBE) formalism was employed for the exchange-correlation function \cite{perdew1996generalized}. For phonon spectra calculations, we used the PHONOPY code to construct the force constants matrices and generate the symmetry information \cite{togo2015first}. The phonon surface states were calculated using the iterative Green’s function method \cite{sancho1985highly} with the tight-binding model Hamiltonian carried out by the WannierTools package \cite{wu2018wanniertools}. The high-throughput calculations were performed on hundreds of candidate compounds from the Inorganic Crystal Structure Database (ICSD), which were selected based on their space groups, to search for ideal 3D Dirac phonon materials.

\section{$\textbf{\emph{k}} \cdot \textbf{\emph{p}}$ Hamiltonian and linear dispersion}
In this section, we derive the low-energy effective Hamiltonian for all the candidate space groups using the $\textbf{\emph{k}} \cdot \textbf{\emph{p}}$ model. The irreducible representations (IRs) are generated by performing the REPRES program, which is carried out by a normal-subgroup induction method \cite{aroyo2006bilbao,elcoro2017double}. All the Hamiltonians and operators are represented by 4$\times$4 matrices, corresponding to four energy levels of Dirac phonons. To reduce the number of parameters, we expand the Hamiltonian with sixteen Dirac $\Gamma$ matrices as
\begin{equation}
\begin{aligned}
H(\textbf{\emph{q}})&=\sum_{\begin{subarray}{1} i=0,x,y,z \\ j=0,x,y,z \end{subarray}} d_{ij}(\textbf{\emph{q}})\Gamma_{ij},\\
\Gamma_{ij}&=\tau_i\otimes\sigma_j,
\end{aligned}
\end{equation}
where $\textbf{\emph{q}}=(q_x,q_y,q_z)$ is the phonon wave vector originating from specific momentum, $d_{ij}(\textbf{\emph{q}})$ are real functions expanded up to the leading order, $\tau_0 (\sigma_0)$ is identity matrix and $\tau_{x,y,z} (\sigma_{x,y,z})$ are Pauli matrices. This Hamiltonian is constrained by the co-little group of the momentum as
\begin{equation}
RH(\textbf{\emph{q}})R^{-1}=H(R\textbf{\emph{q}}),
\end{equation}
where $R$ is any operator in the co-little group, including space group symmetries and time-reversal-related symmetries. For all relevant symmetries, we have summarized their matrix representations and the transformation relations of their reciprocal space coordinates in Table S4 and Table S5. Below, we derive the Hamiltonians for Si (\emph{cI}16) with space group 206 and Nb$_3$Te$_3$As with space group 176 in detail. For the rest cases, we provide the Hamiltonian matrices directly.

\subsection{Si (\emph{cI}16) with space group 206}
At high symmetry point \emph{P} with coordinates $(\pi,\pi,\pi)/a_0$ ($a_0$ is the lattice constant), the operators in the little group  are generated by the rotation symmetry \emph{C}$_{3,111}$ along with two screw symmetries: \emph{S}$_{2x}$ and \emph{S}$_{2y}$. By performing symmetry operations on the lattice, the space coordinates will be transformed from ($x,y,z$) to ($x',y',z'$), thus we can define the key operators as
\begin{align}
&\emph{C}_{3,111}:(x,y,z)\to(z,x,y), \notag \\
&\emph{S}_{2x}:(x,y,z)\to(x,-y,-z+1/2), \notag \\
&\emph{S}_{2y}:(x,y,z)\to(-x+1/2,y,-z), \notag \\
&\emph{S}_{2z}:(x,y,z)\to(-x,-y+1/2,z).
\end{align}
Taking $\emph{S}_{2x}$ and $\emph{S}_{2y}$ as an example, we explore the relations between the operators. Depending on the operating orders, a lattice translation will arise since
\begin{equation}
\emph{S}_{2x}\emph{S}_{2y}=\textbf{\emph{T}}_\emph{z}\emph{S}_{2y}\emph{S}_{2x},
\end{equation}
where $\textbf{\emph{T}}_\emph{z}=\emph{a}_0 \hat{\textbf{\emph{z}}}$. Hence, $\emph{S}_{2x}$ and $\emph{S}_{2y}$ are anti-commutative since $\textbf{\emph{T}}_\emph{z}$ takes the value of $e^{i\textbf{\emph{k}}_P \cdot a_0\hat{\textbf{\emph{z}}}}$, which is -1. Analogously, we can obtain the relations of the operators as below
\begin{equation}
\{S_{2\alpha},S_{2\beta}\}=0,\;\{S_{2\alpha},\mathcal{PT}\}=0,\;[C_{3,111},\mathcal{PT}]=0,\;S_{2\alpha}C_{3,111}S_{2\alpha}^{-1}=S_{2\beta},
\end{equation}
where $\alpha=(x,y,z)$, $\beta=(y,z,x)$ and $\mathcal{PT}$ is the combined symmetry of the inversion symmetry and the time-reversal symmetry. Then, the IRs of the three screw operators can be written as $S_{2x}=-\sigma_y$, $S_{2y}=\sigma_x$  and $S_{2z}=\sigma_z$. Considering the three-fold rotation symmetry, there are three IRs, given by
\begin{align}
&P_1: \;  C_{3,111}= \frac{\sqrt{2}}{2} \begin{bmatrix}
           e^{i\pi/12} & e^{i\pi/12} \\
           e^{-i5\pi/12} & e^{i7\pi/12} \\
          \end{bmatrix},   \notag \\
&P_2: \; C_{3,111}= \frac{\sqrt{2}}{2} \begin{bmatrix}
           e^{i3\pi/4} & e^{i3\pi/4} \\
           e^{i\pi/4} & e^{-i3\pi/4} \\
          \end{bmatrix},   \notag \\
&P_3: \; C_{3,111}= \frac{\sqrt{2}}{2} \begin{bmatrix}
           e^{-i7\pi/12} & e^{-i7\pi/12} \\
           e^{i11\pi/12} & e^{-i\pi/12} \\
          \end{bmatrix}.
\end{align}
Under the effect of time-reversal symmetry, the physical irreducible representations are constructed by $P'=P_1 \oplus P_3$ or $P''=P_2 \oplus P_2$, giving rise to four-fold degeneracy. For the same reason, the IRs of the three screw operators should be written as $S_{2x}=-\sigma_0 \otimes\sigma_y$, $S_{2y}=\sigma_0 \otimes\sigma_x$  and $S_{2z}=\sigma_0 \otimes\sigma_z$. The combined $\mathcal{PT}$ symmetry is represented as $\sigma_y \otimes \sigma_y K$ (\emph{K} is the complex conjugate operator), which satisfies the relations in (5) and squares to 1. By combining equations (1) and (2) with symmetries $S_{2x}$, $S_{2y}$, $C_{3,111}$ and $\mathcal{PT}$, we obtain the simplest form of the Hamiltonian
\begin{equation}
H_P(\textbf{\emph{q}})=A \begin{bmatrix}
           q_z & q_y+iq_x &  &  \\
           q_y-iq_x & -q_z &  &  \\
           &  & -q_z & -q_y-iq_x \\
           &  & -q_y+iq_x & q_z
\end{bmatrix},
\end{equation}
where $\textbf{\emph{q}}=(q_x,q_y,q_z)$ is the phonon wave vector originating from point P and $d_i (\textbf{\emph{q}})$ are expanded up to the lowest order. This is exactly the massless form of Dirac Hamiltonian in three dimensions. Its eigenvalues can be written as
\begin{equation}
\begin{aligned}
&E_{P,1}(\textbf{\emph{q}})=E_{P,2}(\textbf{\emph{q}})=A\sqrt{q_x^2+q_y^2+q_z^2}, \\
&E_{P,3}(\textbf{\emph{q}})=E_{P,4}(\textbf{\emph{q}})=-A\sqrt{q_x^2+q_y^2+q_z^2},
\end{aligned}
\end{equation}
exhibiting four-fold degeneracy at $\textbf{\emph{q}}$=0 and linear dispersion when $\textbf{\emph{q}}\rightarrow$0. Note that the IR of $C_{3,111}$ is chosen as $P''$, and the other IR give similar result.

\subsection{Nb$_3$Te$_3$As with space group 176}
Along the $\Delta$ (0, 0, w) axis, the elements in the little group can be generated by the screw rotation symmetry $S_{6z}$, changing the coordinates of the lattice as
\begin{equation}
S_{6z}: (a,b,c) \to (a-b,a,c+1/2).
\end{equation}
Then an integral lattice translation is generated after six times operations of $S_{6z}$, which introduce an extra phase
\begin{equation}
S_{6z}^6=3\textbf{\emph{T}}_\emph{z}=e^{-ik_z \cdot (3c)},
\end{equation}
where \emph{c} is the lattice constant and $\textbf{\emph{T}}_\emph{z}$ is a unit lattice translation along the \emph{z} direction. Hence, the eigenvalues of $S_{6z}$ can be expressed as
\begin{equation}
E_n=e^{i\pi n/3} \cdot e^{-ik_z c/2}\;(n=0,1,2,3,4,5),
\end{equation}
which can be used to indicate each phonon branch. In the presence of time-reversal symmetry, two-dimensional IRs are composed of (\emph{E}$_1$,\emph{E}$_5$) and (\emph{E}$_2$,\emph{E}$_4$), which are complex conjugate pairs, along the $k_z$ axis. Then the representation of $S_{6z}$ is given by
\begin{equation}
S_{6z}=e^{-ik_z c/2} \begin{bmatrix}
e^{i\frac{\pi}{3}} & & & \\
& e^{i\frac{5\pi}{3}} & & \\
& & e^{i\frac{2\pi}{3}} & \\
& & & e^{i\frac{4\pi}{3}}
\end{bmatrix}.
\end{equation}
The representation of $\mathcal{PT}$ can be written as $\sigma_0\otimes\sigma_x K$ by choosing the same basis functions as $S_{6z}$. Then we expand the Hamiltonian around a specific point on axis $\Delta$ as
\begin{equation}
H_{\Delta}(\textbf{\emph{q}})=d_{yz}(\textbf{\emph{q}})\Gamma_{yz}+\sum_{\begin{subarray}{1} i=0,x,z \\ j=0,x,y \end{subarray}}d_{ij}(\textbf{\emph{q}})\Gamma_{ij},
\end{equation}
which is commutative with the $\mathcal{PT}$ symmetry. Under the constrain of the screw rotation symmetry, the Hamiltonian satisfies
\begin{equation}
S_{6z}H_{\Delta}(\textbf{\emph{q}})S_{6z}^{-1}=H_{\Delta}(S_{6z}\textbf{\emph{q}}).
\end{equation}
Consequently, we can have the final form of the Hamiltonian
\begin{equation}
H_{\Delta}(\textbf{\emph{q}})=Aq_z\Gamma_{z0}+B(q_x\Gamma_{yz}+q_y\Gamma_{x0}).
\end{equation}
Its eigenvalues can be written as
\begin{equation}
\begin{aligned}
&E_{\Delta,1}(\textbf{\emph{q}})=E_{\Delta,2}(\textbf{\emph{q}})=\sqrt{A^2q_z^2+B^2(q_x^2+q_y^2)}, \\
&E_{\Delta,3}(\textbf{\emph{q}})=E_{\Delta,4}(\textbf{\emph{q}})=-\sqrt{A^2q_z^2+B^2(q_x^2+q_y^2)},
\end{aligned}
\end{equation}
exhibiting four-fold degeneracy at $\textbf{\emph{q}}$=0 and linear dispersion when $\textbf{\emph{q}}\rightarrow$0. Due to the screw rotation symmetry, the dispersion of Dirac phonons is isotropic along the directions perpendicular to the axis by considering the lowest order items in the Hamiltonian.

\subsection{Hamiltonians of other space groups}

Firstly, we focus on the $\textbf{\emph{k}}$ vectors which are high symmetry points. For space groups in orthorhombic and tetragonal crystal systems, we define the wave vector as $\textbf{\emph{q}}=(q_x,q_y,q_z)$ since the three basis vectors are orthogonal to each other. The Hamiltonians are given by

\begin{equation}
\begin{aligned}
H_{73(W)}(\textbf{\emph{q}})=E_0 \Gamma_{00}&+q_x(A_1\Gamma_{xz}+A_2\Gamma_{yz}+A_3\Gamma_{zz})\\
&+q_y(B_1\Gamma_{xy}+B_2\Gamma_{yy}+B_3\Gamma_{zy})\\
&+q_z(C_1\Gamma_{xx}+C_2\Gamma_{yx}+C_3\Gamma_{zx}),
\end{aligned}
\end{equation}

\begin{equation}
\begin{aligned}
&H_{52(S)}(\textbf{\emph{q}})=H_{54(U,R)}(\textbf{\emph{q}})=H_{56(U,T)}(\textbf{\emph{q}})=H_{60(T)}(\textbf{\emph{q}})=H_{130(R)}(\textbf{\emph{q}})\\
&=H_{138(R)}(\textbf{\emph{q}})=E_0\Gamma_{00}+Aq_x\Gamma_{zx}+q_y(B\Gamma_{xy}+C\Gamma_{yy})+Dq_z\Gamma_{0y},
\end{aligned}
\end{equation}

\begin{equation}
\begin{aligned}
H_{142(P)}(\textbf{\emph{q}})=E_0 \Gamma_{00}&+A(q_x\Gamma_{xx}+q_y\Gamma_{xy})+B(q_x\Gamma_{yx}+q_y\Gamma_{yy})\\
&+C(q_x\Gamma_{0y}+q_y\Gamma_{0x})+q_z(D\Gamma_{x0}+E\Gamma_{y0}).
\end{aligned}
\end{equation}
Since the leading order of the polynomials is first order, their phonon band structures show linear dispersion along any direction. However, for the below cases, the leading order of $q_x$ and $q_y$ is second order, thus their phonon band structures show quadratic dispersion along $q_x$ and $q_y$ directions.

\begin{equation}
\begin{aligned}
&H_{124(A,Z)}(\textbf{\emph{q}})=H_{126(Z)}(\textbf{\emph{q}})=H_{128(A,Z)}(\textbf{\emph{q}})=H_{130(A,Z)}(\textbf{\emph{q}})=H_{133(A)}(\textbf{\emph{q}})=H_{135(A)}(\textbf{\emph{q}})  \\
&=H_{137(A)}(\textbf{\emph{q}})=E_0 \Gamma_{00}+q_xq_y(A\Gamma_{x0} +B\Gamma_{y0}) +(q_x^2-q_y^2)(C\Gamma_{xz}+D\Gamma_{yz})+q_z(E\Gamma_{xy}+F\Gamma_{yy}),
\end{aligned}
\end{equation}

For space groups in trigonal and hexagonal crystal systems, we define the wave vector as $\textbf{\emph{q}}=(q_1,q_2,q_3)$, where $q_3$ is orthogonal to the $q_1-q_2$ plane. We first give the Hamiltonians for the $\textbf{\emph{k}}$ vectors locating at time-reversal invariant momenta

\begin{equation}
\begin{aligned}
H_{163(A)}(\textbf{\emph{q}})=H_{165(A)}(\textbf{\emph{q}})&=H_{167(T)}(\textbf{\emph{q}})=E_0 \Gamma_{00}+A(q_1-q_2)\Gamma_{zx}\\
&+\frac{A}{\sqrt{3}}(q_1+q_2)\Gamma_{0y}+q_3(E\Gamma_{xy}+F\Gamma_{yy}),
\end{aligned}
\end{equation}

\begin{equation}
\begin{aligned}
H_{176(A)}(\textbf{\emph{q}})=&E_0 \Gamma_{00}+(-\frac{q_1^2}{2}-q_1q_2+q_2^2)(A\Gamma_{x0}+B\Gamma_{y0})+\frac{\sqrt{3}}{2}(q_1^2-2q_1q_2)(B\Gamma_{x0}-A\Gamma_{y0}) \\
&+(-\frac{q_1q_3}{2}+q_2q_3)(C\Gamma_{xz}+D\Gamma_{yz})+\frac{\sqrt{3}}{2}q_1q_3(D\Gamma_{xz}-C\Gamma_{yz})+Eq_3\Gamma_{0y},
\end{aligned}
\end{equation}

\begin{equation}
\begin{aligned}
H_{192(A)}(\textbf{\emph{q}})&=E_0 \Gamma_{00}+\sqrt{3}(q_1^2-q_2^2)(A\Gamma_{xz}-B\Gamma_{yz})\\
&+(q_1^2-4q_1q_2+q2^2)(B\Gamma_{x0}+A\Gamma_{y0})+q_3(C\Gamma_{xy}+D\Gamma_{yy}),
\end{aligned}
\end{equation}

\begin{equation}
\begin{aligned}
H_{193(A)}(\textbf{\emph{q}})=H_{194(A)}(\textbf{\emph{q}})&=E_0 \Gamma_{00}+A(q_1^2-4q_1q_2+q_2^2)\Gamma_{xx}-A\sqrt{3}(q_1^2-q_2^2)\Gamma_{yx}\\
&-B(q_1-q_2)q_3\Gamma_{xy}-\frac{\sqrt{3}}{3}B(q_1+q_2)q_3+Cq_3\Gamma_{zy}.
\end{aligned}
\end{equation}
We can see from the Hamiltonian that, for space groups and $\textbf{\emph{k}}$ vectors 163(A), 165(A) and 167(T), their phonon band structures show linear dispersion along any direction, and for 176(A), 192(A) and 193(A), their phonon band structures show quadratic dispersion along $q_1$ and $q_2$ directions, but show linear dispersion along $q_3$ direction. Then we give the Hamiltonians for the $\textbf{\emph{k}}$ vectors which are not time-reversal invariant momenta. In these cases, $\mathcal{PT}$ symmetry should be used as the anti-unitary operator instead of the time-reversal symmetry.

\begin{equation}
H_{165(H)}(\textbf{\emph{q}})=E_0 \Gamma_{00}+Aq_1\Gamma_{zx}+q_2(B\Gamma_{xy}+C\Gamma_{yy})+Dq_3\Gamma_{0y},
\end{equation}

\begin{equation}
\begin{aligned}
H_{192(H)}(\textbf{\emph{q}})&=E_0 \Gamma_{00}+(q_1-q_2)(A\Gamma_{xz}+B\Gamma_{yz})\\
&+\frac{\sqrt{3}}{3}(q_1+q_2)(A\Gamma_{y0}-B\Gamma_{x0})+q_3(C\Gamma_{xy}+D\Gamma_{yy}),
\end{aligned}
\end{equation}

\begin{equation}
\begin{aligned}
H_{193(H)}(\textbf{\emph{q}})=E_0 \Gamma_{00}+A(q_1+q_2)\Gamma_{zx}-\sqrt{3}B(q_1-q_2)\Gamma_{zy}+q_3(C\Gamma_{xz}+D\Gamma_{yz}).
\end{aligned}
\end{equation}
Around such $\textbf{\emph{k}}$ vectors, their phonon band structures show linear dispersion along any direction.

For space in a cubic crystal system, the three basis vectors are orthogonal to each other, so we define the wave vector as $\textbf{\emph{q}}=(q_x,q_y,q_z)$. The Hamiltonians are given by


\begin{equation}
\begin{aligned}
H_{222(X)}(\textbf{\emph{q}})=E_0 \Gamma_{00}+q_xq_y(A\Gamma_{x0} +B\Gamma_{y0})+(q_x^2-q_y^2)(C\Gamma_{xz}+D\Gamma_{yz})+q_z(E\Gamma_{xy}+F\Gamma_{yy}),
\end{aligned}
\end{equation}

\begin{equation}
\begin{aligned}
H_{222(R)}(\textbf{\emph{q}})=H_{223(R)}(\textbf{\emph{q}})=H_{230(H)}(\textbf{\emph{q}})=E_0 \Gamma_{00}&+A[(q_x^2-q_y^2)\Gamma_{xz}+\frac{\sqrt{3}}{3}(q_x^2+q_y^2-2q_z^2)\Gamma_{y0}]\\
&+B[(q_x^2-q_y^2)\Gamma_{yz}-\frac{\sqrt{3}}{3}(q_x^2+q_y^2-2q_z^2)\Gamma_{x0}],
\end{aligned}
\end{equation}

\begin{equation}
\begin{aligned}
H_{226(L)}(\textbf{\emph{q}})=H_{228(L)}(\textbf{\emph{q}})=E_0 \Gamma_{00}+(q_x+q_y+q_z)(A\Gamma_{xy}+B\Gamma_{yy})+C[(q_x-q_y)\Gamma_{zx}+\frac{\sqrt{3}}{3}(q_x+q_y-2q_z)\Gamma_{0y}],
\end{aligned}
\end{equation}

\begin{equation}
\begin{aligned}
H_{230(P)}(\textbf{\emph{q}})=E_0 \Gamma_{00}&+A(q_x\Gamma_{xy}-q_y\Gamma_{xx}-q_z\Gamma_{xz})\\
&+B(q_x\Gamma_{yy}-q_y\Gamma_{yx}-q_z\Gamma_{yz}),
\end{aligned}
\end{equation}

\begin{equation}
\begin{aligned}
H_{230(P)}(\textbf{\emph{q}})=E_0 \Gamma_{00}&+A(q_x+q_y)(\Gamma_{zx}-\Gamma_{zy})+(\sqrt{3}+1)B(q_x\Gamma_{yz}-q_x\Gamma_{x0}+q_y\Gamma_{xz}-q_y\Gamma_{y0})\\
&-A(q_x-q_y)(\Gamma_{0x}+\Gamma_{0y})+(\sqrt{3}-1)B(q_x\Gamma_{xz}+q_x\Gamma_{y0}+q_y\Gamma_{yz}+q_y\Gamma_{x0})\\
&+2Aq_z\Gamma_{0z}+2Bq_z(\Gamma_{xx}+\Gamma_{xy}-\Gamma_{yx}-\Gamma_{yy}),
\end{aligned}
\end{equation}
For space groups and $\textbf{\emph{k}}$ vectors 226(L), 228(L) and 230(P), their phonon band structures show linear dispersion along any direction. Note that there two types of IRs at 230(P). For 222(X), its phonon band structures show quadratic dispersion along $q_x$ and $q_y$ directions and linear dispersion along $q_z$ direction. For 222(R), 223(R) and 230(H), their phonon band structures show quadratic dispersion along any direction.

Next, we give the Hamiltonian for the high symmetry lines. Since the leading order is always first order, their phonon band structures show linear dispersion along any direction.

\begin{equation}
\begin{aligned}
&H_{55(Q)}(\textbf{\emph{q}})=H_{56(Q)}(\textbf{\emph{q}})=H_{58(Q)}(\textbf{\emph{q}})=H_{59(Q)}(\textbf{\emph{q}})=H_{62(P)}(\textbf{\emph{q}})=H_{62(E)}(\textbf{\emph{q}})\\
&=E_0 \Gamma_{00}+q_x(A_1\Gamma_{xx}+A_2\Gamma_{xy})+q_y(B_1\Gamma_{x0}+B_2\Gamma_{yz})+(C+Dq_z)\Gamma_{z0},
\end{aligned}
\end{equation}

\begin{equation}
\begin{aligned}
H_{175(\Delta)}(\textbf{\emph{q}})&=E_0 \Gamma_{00}+[A(2q_1-q_2)-\sqrt{3}Bq_2]\Gamma_{xx}\\
&+[\sqrt{3}Aq_2+B(2q_1-q_2)]\Gamma_{xy}+(C+Dq_3)\Gamma_{z0},
\end{aligned}
\end{equation}

\begin{equation}
\begin{aligned}
&H_{191(\Delta)}(\textbf{\emph{q}})=H_{192(\Delta)}(\textbf{\emph{q}})=H_{193(\Delta)}(\textbf{\emph{q}})=H_{194(\Delta)}(\textbf{\emph{q}})\\
&=E_0 \Gamma_{00}+A(q_1-q_2)\Gamma_{xx}+\frac{\sqrt{3}}{3}A(q_1+q_2)\Gamma_{xy}+(B+Cq_3)\Gamma_{z0},
\end{aligned}
\end{equation}

\section{Nodal-line in SiP$_2$ with space group 205}
In this section, we give an example to show the formation mechanism of nodal-line induced by accidental degeneracy along the high symmetry line. In Fig. S1 (a), we show the first Brillouin zone of SiP$_2$, which belongs to the space group \emph{Pa}$\overline{3}$ (No. 205). Then we focus on its phonon band structure near 5.7 THz. As shown in the dashed boxes of Fig. S1 (c) and (d), a band cross with four-fold degeneracy and linear dispersion can be found along the Z path. However, is this a Dirac phonon with at least two bands completely separated everywhere around it? To verify this point, one should check the phonon band structure in the high symmetry planes that pass through the Z path. Away from the Z path,the band gap is opened along the paths B$_1$-B$_2$ and B$_3$-B$_4$ [as shown in Fig. S1 (c)] and bands split into four single bands with four degenerate points along the paths A$_1$-A$_2$ and A$_3$-A$_4$ [as shown in Fig. S1 (d)]. Since the $\emph{k}$-paths are randomly selected, nodal-line must exist in plane A. Such features are coincident with the compatibility relations that we give in Fig. S1 (b). To provide an intuitive picture, we label the phonon branches with corresponding IRs in Fig. S1 (c) and (d). In plane A, band crossings are formed by phonon branches carrying different IRs and thus can not be opened, implying the existence of nodal-line instead of Dirac phonon.

\begin{figure}[bp]
	\centering
	\includegraphics[scale=0.8]{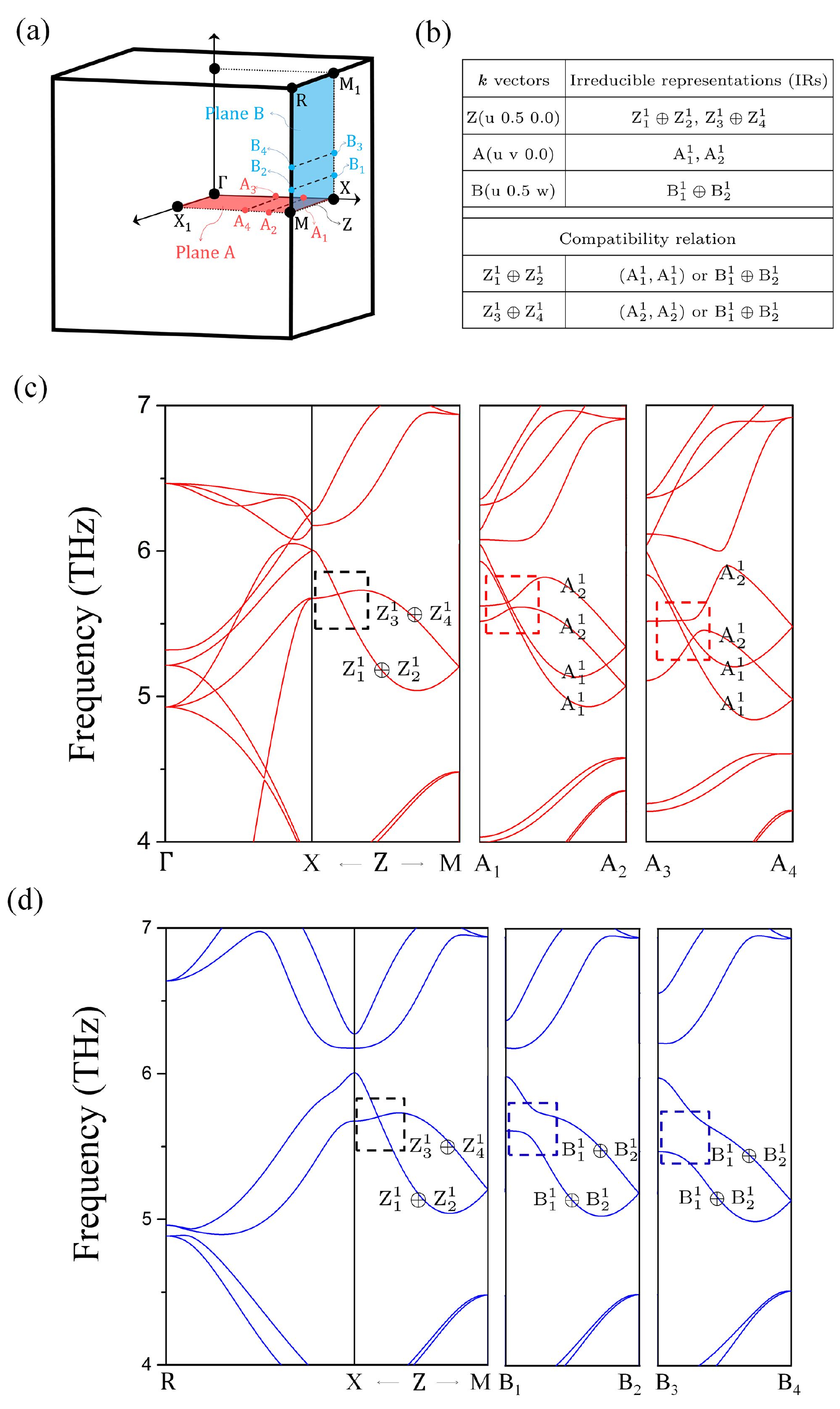}
	\renewcommand{\thefigure}{S\arabic{figure}}
     \caption{(a) The first Brillouin zone of space group 205. (b) IRs and compatibility relations of Z, A and B. (c) and (d) Phonon band structures of SiP$_2$ with labeled IRs.}
\end{figure}

\section{Other candidate materials}
In Fig. S2, we plot the first Brillouin zone for the candidate space groups and mark the possible positions for three-dimensional (3D) Dirac phonons. For each case, we performed high-throughput calculations to search for realistic 3D Dirac phonon materials. The phonon spectra are shown in Fig. S3.

\begin{figure}[!hp]
	\centering
	\includegraphics[scale=0.056]{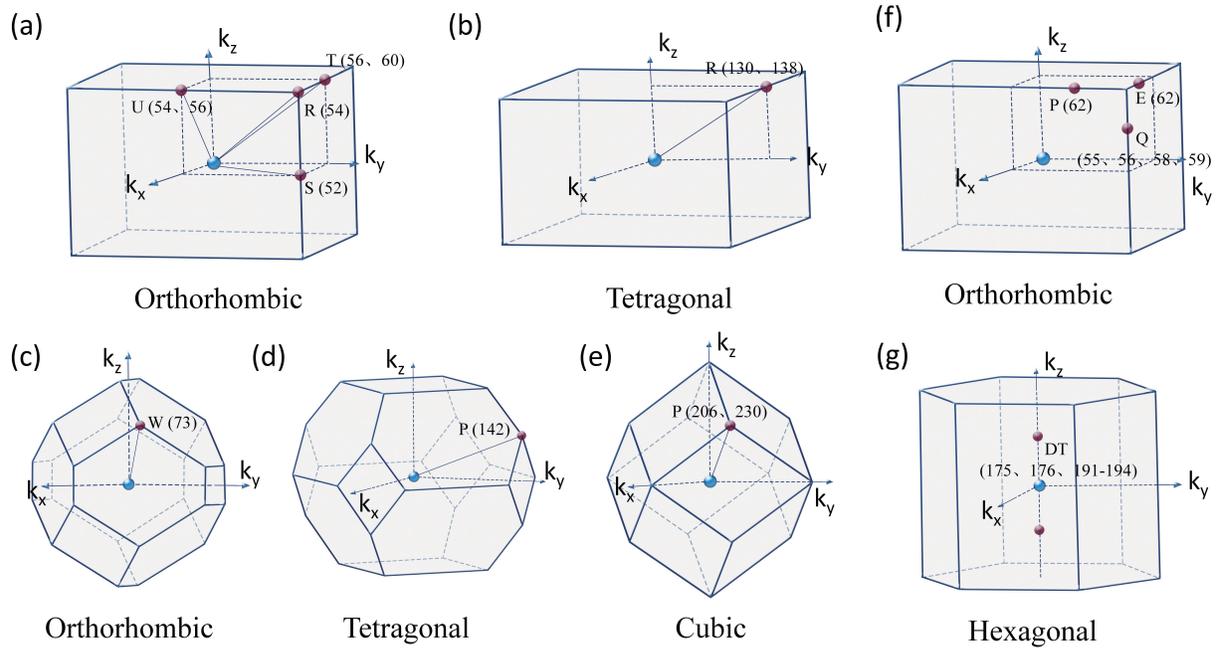}
	\renewcommand{\thefigure}{S\arabic{figure}}
     \caption{(a)-(f) The first Brillouin zone for the candidate space groups with 3D Dirac phonons at the high symmetry points. (g)-(h) The first Brillouin zone for the candidate space groups with 3D Dirac phonons along the high symmetry lines.}
\end{figure}

\clearpage

\begin{figure}
	\centering
	\includegraphics[scale=0.048]{FIG_S3_a.pdf}
\end{figure}

\begin{figure}
	\centering
	\includegraphics[scale=0.048]{FIG_S3_b.pdf}
\end{figure}

\begin{figure}
	\centering
	\includegraphics[scale=0.048]{FIG_S3_c.pdf}
\end{figure}

\begin{figure}
	\centering
	\includegraphics[scale=0.048]{FIG_S3_d.pdf}
\end{figure}

\begin{figure}
	\centering
	\includegraphics[scale=0.048]{FIG_S3_e.pdf}
\end{figure}

\begin{figure}
	\centering
	\includegraphics[scale=0.048]{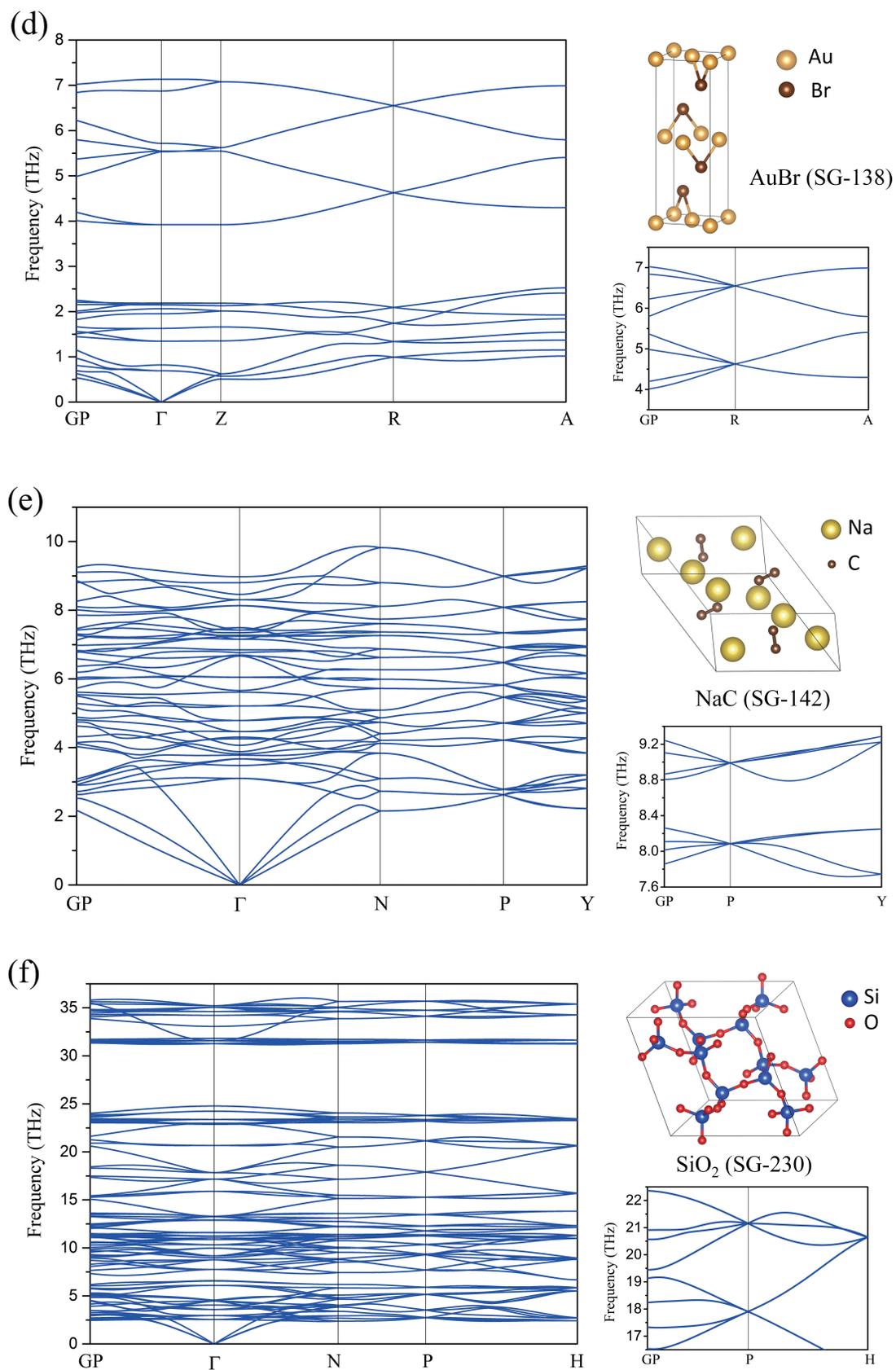}
	\renewcommand{\thefigure}{S\arabic{figure}}
     \caption{Crystal structures and phonon spectra of candidate 3D Dirac phonon materials.}
\end{figure}

\clearpage

\section{Materials with quadratic dispersion or nodal-lines}

\begin{figure}[!hbp]
	\centering
	\includegraphics[scale=0.16]{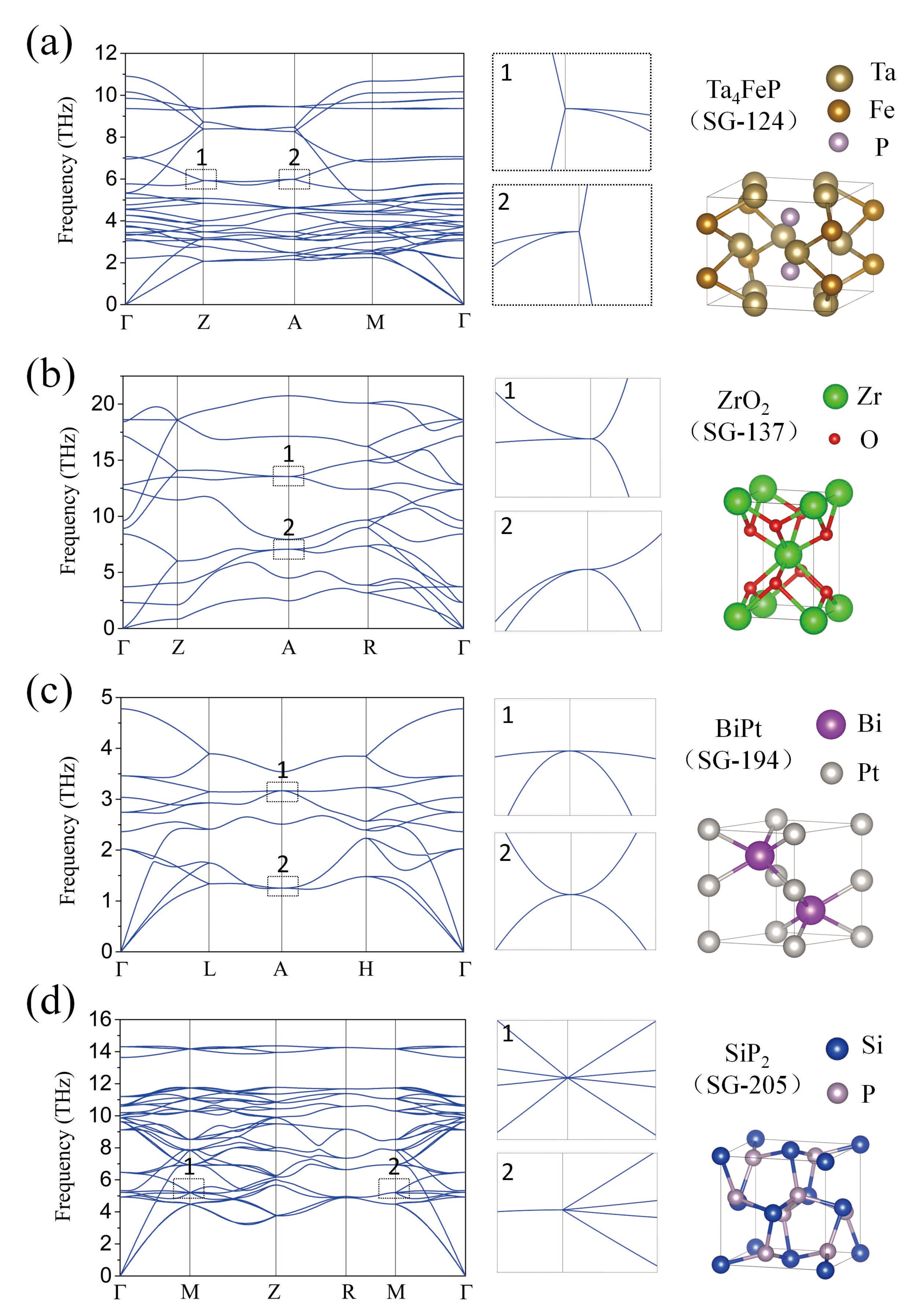}
	\renewcommand{\thefigure}{S\arabic{figure}}
     \caption{Crystal structures and phonon spectra of materials that (a)-(c) have four-fold band degeneracy at high symmetry points but show quadratic dispersion and (d) have four-fold degenerate nodal-line passing through high symmetry points.}
\end{figure}

\clearpage

\section{Appendix}

\begin{table}[!htbp]
\footnotesize
\renewcommand{\thetable}{S\arabic{table}}
\renewcommand\arraystretch{2.6}
\setlength{\tabcolsep}{6.5mm}
\begin{tabular}{c|c|c|c}
\toprule[0.1em]
\hline
\hline
 Space group & HSP & IR & Only 4D IR\\
\hline
\hline
           52             &   S(0.5 0.5 0.0)   &   S$_1^2\oplus $S$_2^2$ 										         &	yes	\\ \cline{1-4}
   \multirow{2}{*}{54}    &   R(0.5 0.5 0.5)   &   R$_1^2\oplus $R$_2^2$ 										         &	yes	\\ \cline{2-4}
                          &   U(0.5 0.0 0.5)   &   U$_1^2\oplus $U$_2^2$ 										         &	yes	\\ \cline{1-4}
   \multirow{2}{*}{56}    &   T(0.0 0.5 0.5)   &   T$_1^2\oplus $T$_2^2$ 										         &	yes	\\ \cline{2-4}
                          &   U(0.5 0.0 0.5)   &   U$_1^2\oplus $U$_2^2$ 										         &	yes	\\ \cline{1-4}
           60             &   T(0.0 0.5 0.5)   &   T$_1^2\oplus $T$_2^2$ 										         &	yes	\\ \cline{1-4}
           73             &   W(0.5 0.5 0.5)   &   W$_1^2\oplus $W$_1^2$ 										         &	yes	\\
\end{tabular}
\caption{High symmetry points (HSPs) and their corresponding space groups that can host Dirac phonons. ``S$_n^m$'' indicates the $n$-th IR at S point with $m$ dimension. The symbol ``$\oplus$'' represents the direct sum of two sets of IRs combined by the time-reversal symmetry. ``4D'' is the abbreviation of ``four-dimensional''.}
\end{table}

\clearpage

\begin{table}[!htbp]
\footnotesize
\renewcommand{\thetable}{S\arabic{table}}
\renewcommand\arraystretch{2.3}
\setlength{\tabcolsep}{6.5mm}
\begin{tabular}{c|c|c|c}
\toprule[0.1em]
\hline
\hline
Space group & HSP & IR & Remarks \\
\hline                      	
   \multirow{2}{*}{124}   &   A(0.5 0.5 0.5)   &   A$_1^2,$ A$_2^2,$ A$_3^2\oplus $A$_4^2$ 											&	QD		\\ \cline{2-4}
                          &   Z(0.0 0.0 0.5)   &   Z$_1^2,$ Z$_2^2,$ Z$_3^2\oplus $Z$_4^2$ 											&	QD		\\ \cline{1-4}
           126            &   Z(0.0 0.0 0.5)   &   Z$_1^2,$ Z$_2^2,$ Z$_3^2\oplus $Z$_4^2$ 											&	QD		\\ \cline{1-4}
   \multirow{2}{*}{128}   &   A(0.5 0.5 0.5)   &   A$_1^2,$ A$_2^2,$ A$_3^2\oplus $A$_4^2$ 											&	QD		\\ \cline{2-4}
                          &   Z(0.0 0.0 0.5)   &   Z$_1^2,$ Z$_2^2,$ Z$_3^2\oplus $Z$_4^2$ 											&	QD		\\ \cline{1-4}
   \multirow{2}{*}{130}   &   A(0.5 0.5 0.5)   &   A$_1^2\oplus $A$_2^2,$ A$_3^2\oplus $A$_4^2$ 									&	QD		\\ \cline{2-4}
                          &   Z(0.0 0.0 0.5)   &   Z$_1^2,$ Z$_2^2,$ Z$_3^2\oplus $Z$_4^2$ 		 									&	QD		\\
\end{tabular}
\caption{High symmetry points (HSPs) and their corresponding space groups that involve four-dimensional IRs but can not host Dirac phonons. ``QD'' means the phonon band structure with quadratic dispersion along some directions. ``NL'' means nodal-line with four-fold band degeneracy passing through such high symmetry point.}
\end{table}

\clearpage

\begin{table}[!htbp]
\scriptsize
\renewcommand{\thetable}{S\arabic{table}}
\renewcommand\arraystretch{2.0}
\setlength{\tabcolsep}{4.0mm}
\begin{tabular}{c|c|c|c|c|c}
\toprule[0.1em]
\hline
\hline
Space group & Line & Plane & \multicolumn{2}{|c|}{Compatibility relation} & Node type \\
\hline
\multirow{4}{*}{51}   			&  \multirow{2}{*}{G(0.5 0.0 w)}    		&   \multirow{2}{*}{M(u 0.0 w) or L(0.5 v w)}	&   $\rm G_1^1\oplus G_2^1$   	&  $\rm (M_1^1,M_1^1)$ or $\rm L_1^1\oplus L_2^1$    	&  \multirow{2}{*}{NL}     \\ \cline{4-5}
                      			&                                   		&                              	              	&   $\rm G_3^1\oplus G_4^1$   	&  $\rm (M_2^1,M_2^1)$ or $\rm L_1^1\oplus L_2^1$    	&                          \\ \cline{2-6}
  				   			&  \multirow{2}{*}{Q(0.5 0.5 w)}    		&   \multirow{2}{*}{N(u 0.5 w) or L(0.5 v w)} 	&   $\rm Q_1^1\oplus Q_2^1$   	&  $\rm (N_1^1,N_1^1)$ or $\rm L_1^1\oplus L_2^1$    	&  \multirow{2}{*}{NL}     \\ \cline{4-5}
                      			&                                   		&                              	                &   $\rm Q_3^1\oplus Q_4^1$   	&  $\rm (N_2^1,N_2^1)$ or $\rm L_1^1\oplus L_2^1$    	&                          \\ \cline{1-6}
\multirow{4}{*}{52}   			&  \multirow{2}{*}{Q(0.5 0.5 w)}    		&   \multirow{2}{*}{L(0.5 v w) or N(u 0.5 w)}	&   $\rm G_1^1\oplus G_2^1$   	&  $\rm (L_1^1,L_1^1)$ or $\rm N_1^1\oplus N_2^1$    	&  \multirow{2}{*}{NL}     \\ \cline{4-5}
                      			&                                   		&                              	              	&   $\rm G_3^1\oplus G_4^1$   	&  $\rm (L_2^1,L_2^1)$ or $\rm N_1^1\oplus N_2^1$    	&                          \\ \cline{2-6}
 				   			&  \multirow{2}{*}{C(u 0.5 0)}    		&   \multirow{2}{*}{V(u v 0.0) or N(u 0.5 w)} 	&   $\rm Q_1^1\oplus Q_2^1$   	&  $\rm (V_1^1,V_1^1)$ or $\rm N_1^1\oplus N_2^1$    	&  \multirow{2}{*}{NL}     \\
\end{tabular}
\end{table}

\clearpage

\begin{table}[!htbp]
\scriptsize
\renewcommand{\thetable}{S\arabic{table}}
\renewcommand\arraystretch{2.0}
\setlength{\tabcolsep}{4.0mm}
\begin{tabular}{c|c|c|c|c|c}
\toprule[0.1em]
\hline
\hline
Space group & Line & Plane & \multicolumn{2}{|c|}{Compatibility relation} & Node type \\
\hline
\multirow{6}{*}{58}   			&  \multirow{2}{*}{Q(0.5 0.5 w)}    		&   \multirow{2}{*}{L(0.5 v w) or N(u 0.5 w)}	&   $\rm Q_1^1\oplus Q_2^1$   	&  $\rm L_1^1\oplus L_2^1$ or $\rm N_1^1\oplus N_2^1$    	&  \multirow{2}{*}{DP}     \\ \cline{4-5}
                      			&                                   		&                              	              	&   $\rm Q_3^1\oplus Q_4^1$   	&  $\rm L_1^1\oplus L_2^1$ or $\rm N_1^1\oplus N_2^1$    	&                          \\ \cline{2-6}
  				   			&  \multirow{2}{*}{D(0.5 v 0.0)}    		&   \multirow{2}{*}{V(u v 0.0) or L(0.5 v w)} 	&   $\rm D_1^1\oplus D_2^1$   	&  $\rm (V_1^1,V_1^1)$ or $\rm L_1^1\oplus L_2^1$    		&  \multirow{2}{*}{NL}     \\ \cline{4-5}
                      			&                                   		&                              	                &   $\rm D_3^1\oplus D_4^1$   	&  $\rm (V_2^1,V_2^1)$ or $\rm L_1^1\oplus L_2^1$    		&                          \\ \cline{2-6}
      			   			&  \multirow{2}{*}{C(u 0.5 0.0)}    		&   \multirow{2}{*}{V(u v 0.0) or N(u 0.5 w)}	&   $\rm C_1^1\oplus C_2^1$   	&  $\rm (V_1^1,V_1^1)$ or $\rm N_1^1\oplus N_2^1$   	 	&  \multirow{2}{*}{NL}     \\ \cline{4-5}
                      			&                                   		&                              	              	&   $\rm C_3^1\oplus C_4^1$   	&  $\rm (V_2^1,V_2^1)$ or $\rm N_1^1\oplus N_2^1$   	 	&                          \\ \cline{1-6}
\multirow{6}{*}{59}   			&  \multirow{2}{*}{G(0.5 0.0 w)}    		&   \multirow{2}{*}{M(u 0.0 w) or L(0.5 v w)}	&   $\rm G_1^1\oplus G_2^1$   	&  $\rm (M_1^1,M_1^1)$ or $\rm L_1^1\oplus L_2^1$  	  	&  \multirow{2}{*}{NL}     \\
\end{tabular}
\end{table}

\clearpage

\begin{table}[!htbp]
\scriptsize
\renewcommand{\thetable}{S\arabic{table}}
\renewcommand\arraystretch{2.0}
\setlength{\tabcolsep}{4.0mm}
\begin{tabular}{c|c|c|c|c|c}
\toprule[0.1em]
\hline
\hline
Space group & Line & Plane & \multicolumn{2}{|c|}{Compatibility relation} & Node type \\
\hline
\multirow{4}{*}{128}   			&  \multirow{2}{*}{Y(u 0.5 0.0)}    		&   \multirow{2}{*}{D(u v 0.0) or F(u 0.5 w)} 	&   $\rm Y_1^1\oplus Y_2^1$   	&  $\rm (D_1^1,D_1^1)$ or $\rm F_1^1\oplus F_2^1$    	&  \multirow{2}{*}{NL}     \\ \cline{4-5}
                      			&                                   		&                              	                &   $\rm Y_3^1\oplus Y_4^1$   	&  $\rm (D_2^1,D_2^1)$ or $\rm F_1^1\oplus F_2^1$    	&                          \\ \cline{2-6}
      			   			&  \multirow{3}{*}{V(0.5 0.5 w)}    		&   \multirow{3}{*}{C(u u w) or F(u 0.5 w)}	&   $\rm V_1^1\oplus V_2^1$   	&  $\rm (C_1^1,C_1^1)$ or $\rm F_1^1\oplus F_2^1$    	&  \multirow{3}{*}{NL}     \\ \cline{4-5}
                      			&                                   		&                              	              	&   $\rm V_3^1\oplus V_4^1$   	&  $\rm (C_2^1,C_2^1)$ or $\rm F_1^1\oplus F_2^1$    	&                          \\ \cline{4-5}
                      			&                                   		&                              	              	&   $\rm V_5^2$   			&  $\rm (C_1^1,C_2^1)$ or $\rm F_1^1\oplus F_2^1$    	&                          \\ \cline{1-6}
\multirow{4}{*}{129}   			&  \multirow{2}{*}{W(0.0 0.5 w)}    		&   \multirow{2}{*}{B(0.0 v w) or F(u 0.5 w)}	&   $\rm W_1^1\oplus W_2^1$   	&  $\rm (B_1^1,B_1^1)$ or $\rm F_1^1\oplus F_2^1$    	&  \multirow{2}{*}{NL}     \\ \cline{4-5}
                      			&                                   		&                              	              	&   $\rm W_3^1\oplus W_4^1$   	&  $\rm (B_2^1,B_2^1)$ or $\rm F_1^1\oplus F_2^1$    	&                          \\
\end{tabular}
\end{table}

\clearpage

\begin{table}[!htbp]
\scriptsize
\renewcommand{\thetable}{S\arabic{table}}
\renewcommand\arraystretch{2.0}
\setlength{\tabcolsep}{4.0mm}
\begin{tabular}{c|c|c|c|c|c}
\toprule[0.1em]
\hline
\hline
Space group & Line & Plane & \multicolumn{2}{|c|}{Compatibility relation} & Node type \\
\hline
\multirow{2}{*}{191}   			&  \multirow{2}{*}{$\Delta$(0.0 0.0 w)}	&   \multirow{2}{*}{D(u 0.0 w) or C(u u w)}	&   $\rm \Delta_5^2$   		&  $\rm (D_1^1,D_2^1)$ or $\rm (C_1^1,C_2^1)$    	&  \multirow{2}{*}{DP}     \\ \cline{4-5}
                      			&                                   		&                              	              	&   $\rm \Delta_6^2$   		&  $\rm (D_1^1,D_2^1)$ or $\rm (C_1^1,C_2^1)$    	&                          \\ \cline{1-6}
\multirow{2}{*}{192}   			&  \multirow{2}{*}{$\Delta$(0.0 0.0 w)}	&   \multirow{2}{*}{D(u 0.0 w) or C(u u w)}	&   $\rm \Delta_5^2$   		&  $\rm (D_1^1,D_2^1)$ or $\rm (C_1^1,C_2^1)$    	&  \multirow{2}{*}{DP}     \\ \cline{4-5}
                      			&                                   		&                              	              	&   $\rm \Delta_6^2$   		&  $\rm (D_1^1,D_2^1)$ or $\rm (C_1^1,C_2^1)$    	&                          \\ \cline{1-6}
\multirow{4}{*}{193}   			&  \multirow{2}{*}{Q(u u 0.5)}    		&   \multirow{2}{*}{C(u u w) or E(u v 0.5)}	&   $\rm Q_1^1\oplus Q_2^1$   	&  $\rm (C_1^1,C_1^1)$ or $\rm E_1^1\oplus E_2^1$    	&  \multirow{2}{*}{NL}     \\ \cline{4-5}
                      			&                                   		&                              	              	&   $\rm Q_3^1\oplus Q_4^1$   	&  $\rm (C_2^1,C_2^1)$ or $\rm E_1^1\oplus E_2^1$    	&                          \\ \cline{2-6}
				   			&  \multirow{2}{*}{$\Delta$(0.0 0.0 w)}	&   \multirow{2}{*}{D(u 0.0 w) or C(u u w)}	&   $\rm \Delta_5^2$   		&  $\rm (D_1^1,D_2^1)$ or $\rm (C_1^1,C_2^1)$    	&  \multirow{2}{*}{DP}     \\
\end{tabular}
\caption{High symmetry lines (HSLs) and their corresponding space groups that have two or more sets of two-dimensional IRs and their compatibility relations with the high symmetry planes. ``NL'' means nodal-line that must exists in the high symmetry plane passing through the HSL. ``DP'' means Dirac phonon that can exist along the HSL.}
\end{table}

\clearpage

\begin{table}[!htbp]
\scriptsize
\renewcommand{\thetable}{S\arabic{table}}
\renewcommand\arraystretch{1.8}
\setlength{\tabcolsep}{4.5mm}
\begin{tabular}{c|c|cccc|c}
\toprule[0.1em]
\hline
\hline
HSPs		&  \textbf{\emph{T}}($mnl$)	&	\multicolumn{4}{|c|}{Key operators}	&  $\mathcal{T}$ or	$\mathcal{PT}$	\\
\hline

\multirow{3}{*}{52(S)}  &  \multirow{3}{*}{$\{1,-1\}$}    &  $\{I|000\}$				  &  $\{C_{2x}|0\frac{1}{2}\frac{1}{2}\}$   &  	$\{C_{2z}|\frac{1}{2}00\}$     		    &	&  \multirow{3}{*}{$\mathcal{T}=\sigma_x\otimes \sigma_0 K$}	\\
                        &                                 &  $(-q_x,-q_y,-q_z)$			  &  $(q_x,-q_y,-q_z)$                      &  	$(-q_x,-q_y,q_z)$     		            &	&  	\\
                        &  							      &  $\sigma_0\otimes \sigma_z$   &  $\sigma_0\otimes \sigma_x$   			&   $\sigma_z\otimes \sigma_y$   	        &	&												\\ \cline{1-7}

\multirow{3}{*}{54(U)}	&  \multirow{3}{*}{$\{1,-1\}$}    &  $\{I|000\}$				  &  $\{C_{2z}|\frac{1}{2}00\}$   			&  	$\{C_{2y}|00\frac{1}{2}\}$     		    &	&  \multirow{3}{*}{$\mathcal{T}=\sigma_x\otimes \sigma_0 K$}	\\
                        &                                 &  $(-q_x,-q_y,-q_z)$			  &  $(-q_x,-q_y,q_z)$                      &  	$(-q_x,q_y,-q_z)$     		            &	&  	\\
					    &  							      &  $\sigma_0\otimes \sigma_z$   &  $\sigma_0\otimes \sigma_x$   			&   $\sigma_z\otimes \sigma_y$   	        &	&  												\\ \cline{1-7}

\multirow{3}{*}{54(R)}	&  \multirow{3}{*}{$\{1,-1\}$} 	  &  $\{I|000\}$				  &  $\{C_{2z}|\frac{1}{2}00\}$   			&  	$\{C_{2y}|00\frac{1}{2}\}$     		    &	&  \multirow{3}{*}{$\mathcal{T}=\sigma_x\otimes \sigma_0 K$}	\\
\end{tabular}
\end{table}

\clearpage

\begin{table}[!htbp]
\scriptsize
\renewcommand{\thetable}{S\arabic{table}}
\renewcommand\arraystretch{1.8}
\setlength{\tabcolsep}{1.5mm}
\begin{tabular}{c|c|cccc|c}
\toprule[0.1em]
\hline
\hline
HSPs		&  \textbf{\emph{T}}($mnl$)	&	\multicolumn{4}{|c|}{Key operators}	&  $\mathcal{T}$ or	$\mathcal{PT}$	\\
\hline

\multirow{4}{*}{135(A)}	&  \multirow{4}{*}{$\{1,-1\}$}	 &  $\{I|000\}$                  &    $\{M_{y}|0\frac{1}{2}0\}$   		&    $\{C_{4z}^+|\frac{1}{2}0\frac{1}{2}\}$ &	&  \multirow{4}{*}{$\mathcal{T}=\sigma_x\otimes \sigma_0 K$}	\\ 
                        &                                &  $(-q_x,-q_y,-q_z)$			 &    $(q_x,-q_y,q_z)$                  &  	 $(-q_y,q_x,q_z)$     		            &	&  	                                            \\
					    &  							     &  $\sigma_0\otimes \sigma_z$   &    $i\sigma_z\otimes \sigma_z$ 		&    $i\sigma_z\otimes \sigma_x$  			&	&												\\
					    &  							     &  $\sigma_0\otimes \sigma_z$   &    $i\sigma_z\otimes \sigma_0$ 		&    $-\sigma_z\otimes \sigma_y$  			&	&												\\ \cline{1-7}

\multirow{3}{*}{137(A)}	&  \multirow{3}{*}{$\{1,-1\}$}	 &  $\{I|000\}$                  &    $\{M_{y}|0\frac{1}{2}0\}$   		&    $\{C_{4z}^+|\frac{1}{2}0\frac{1}{2}\}$ &	&  \multirow{3}{*}{$\mathcal{T}=\sigma_x\otimes \sigma_0 K$}	\\ 
                        &                                &  $(-q_x,-q_y,-q_z)$			 &    $(q_x,-q_y,q_z)$                  &  	 $(-q_y,q_x,q_z)$     		            &	&  	\\
					    &  							     &  $\sigma_0\otimes \sigma_z$   &    $\sigma_z\otimes \sigma_y$ 		&    $-i\sigma_z\otimes \sigma_0$  			&	&												\\
\end{tabular}
\end{table}

\clearpage

\begin{table}[!htbp]
\scriptsize
\renewcommand{\thetable}{S\arabic{table}}
\renewcommand\arraystretch{1.8}
\setlength{\tabcolsep}{3.0mm}
\begin{tabular}{c|c|cccc|c}
\toprule[0.1em]
\hline
\hline
HSPs		&  \textbf{\emph{T}}($mnl$)	&	\multicolumn{4}{|c|}{Key operators}	&  $\mathcal{T}$ or	$\mathcal{PT}$	\\
\hline

\multirow{3}{*}{222(R)} & \multirow{3}{*}{$\{1,-1\}$} & $\{I|000\}$ & $\{C_{2,110}|00\frac{1}{2}\}$ & $\{C_{3,111}^+|000\}$ & $\{C_{2y}|\frac{1}{2}0\frac{1}{2}\}$ \& $\{C_{2z}|\frac{1}{2}\frac{1}{2}0\}$ & \multirow{3}{*}{$\mathcal{T}=\sigma_x\otimes \sigma_0 K$}	\\
            &           &  $(-q_x,-q_y,-q_z)$			&    $(q_y,q_x,-q_z)$             &  	 $(q_z,q_x,q_y)$     & $(-q_x,q_y,-q_z)$ \& $(-q_x,-q_y,q_z)$    &  	\\
			&    		&  $\sigma_0\otimes \sigma_z$	&     $\sigma_0\otimes \sigma_x$  &    $e^{i2\pi(\sigma_z\otimes \sigma_z)/3}$ 	&	$\sigma_0\otimes \sigma_0$		&		\\ \cline{1-7}

\multirow{3}{*}{222(X)}	&  \multirow{3}{*}{$\{1,-1\}$}	 &  $\{I|000\}$                  &    $\{C_{2x}|0\frac{1}{2}\frac{1}{2}\}$  &    $\{C_{4y}^+|00\frac{1}{2}\}$     	&	&  \multirow{3}{*}{$\mathcal{T}=\sigma_x\otimes \sigma_0 K$}	\\
                        &                                &  $(-q_x,-q_y,-q_z)$			 &    $(q_x,-q_y,-q_z)$                     &  	 $(q_z,q_y,-q_x)$     		        &	&  	\\
					    &  							     &  $\sigma_0\otimes \sigma_z$   &    $\sigma_z\otimes \sigma_y$ 			&    $-i\sigma_z\otimes \sigma_z$  		&	&												\\
\end{tabular}
\caption{IRs and operations of the key space group symmetries and time-reversal-related symmetries for all the centrosymmetric spaces groups that can host Dirac phonons at the high symmetry points (HSPs). \textbf{\emph{T}}($mnl$)=$m$\textbf{\emph{a}}+$n$\textbf{\emph{b}}+$l$\textbf{\emph{c}} are translation operators.}
\end{table}

*$\tau^1=K.$

\clearpage

\begin{table}[!htbp]
\scriptsize
\renewcommand{\thetable}{S\arabic{table}}
\renewcommand\arraystretch{2.0}
\setlength{\tabcolsep}{5.5mm}
\begin{tabular}{c|c|cc|c}
\toprule[0.1em]
\hline
\hline
HSLs						&  \textbf{\emph{T}}($mnl$)					&									\multicolumn{2}{|c|}{Key operators}								&  $\mathcal{PT}$							\\
\hline
\multirow{3}{*}{55(Q)}		&  \multirow{3}{*}{$\{e^{i\theta},-e^{i\theta}\}$} 	&  $\{C_{2z}|000\}$					&  $\{M_x|\frac{1}{2}\frac{1}{2}0\}$  		&  \multirow{3}{*}{$\sigma_0\otimes \sigma_x K$}	\\ 
                            &                                                   &  $(-q_x,-q_y,q_z)$			    &  $(-q_x,q_y,q_z)$                         &                                       \\
						    &  										            &  $\sigma_z\otimes \sigma_0$		&  $i\sigma_0\otimes \sigma_z$				&										\\ 
\hline
\multirow{3}{*}{56(Q)}		&  \multirow{3}{*}{$\{e^{i\theta},-e^{i\theta}\}$} 	&  $\{C_{2z}|\frac{1}{2}\frac{1}{2}0\}$	 &  $\{M_x|\frac{1}{2}0\frac{1}{2}\}$  		 &  \multirow{3}{*}{$\sigma_0\otimes \sigma_x K$}	\\ 
                            &                                                   &  $(-q_x,-q_y,q_z)$			         &  $(-q_x,q_y,q_z)$                         &                                       \\
						    &  										            &  $\sigma_z\otimes \sigma_0$			 &  $\sigma_0\otimes \sigma_z$           	 & 										 \\ 
\hline
\multirow{3}{*}{58(Q)}		&  \multirow{3}{*}{$\{e^{i\theta},-e^{i\theta}\}$} 	&  $\{C_{2z}|000\}$					&  $\{M_x|\frac{1}{2}\frac{1}{2}\frac{1}{2}\}$   &  \multirow{3}{*}{$\sigma_0\otimes \sigma_x K$}	\\
\end{tabular}
\caption{IRs of the key space group symmetries and time-reversal-related symmetries for all the centrosymmetric spaces groups that can host Dirac phonons at the high symmetry lines (HSLs).\textbf{\emph{T}}($mnl$)=$m$\textbf{\emph{a}}+$n$\textbf{\emph{b}}+$l$\textbf{\emph{c}} are translation operators.}
\end{table}

\clearpage

\end{document}